\documentclass[modern]{aastex62}

\begin{document}
\shorttitle{Observations of SN 2013gs}

\shortauthors{Zhang, T. et al.}

\title{Observations of A Fast-Expanding and UV-Bright Type Ia Supernova SN 2013gs}

\author{Tianmeng Zhang}
\email{zhangtm@nao.cas.cn}
\affil{Key Laboratory of Optical Astronomy, National Astronomical Observatories, Chinese Academy of Sciences, Beijing 100012, China}
\affil{School of Astronomy and Space Science, University of Chinese Academy of Sciences}

\author{Xiaofeng Wang}
\affil{Physics Department and Tsinghua Center for Astrophysics (THCA), Tsinghua University, Beijing, 100084, China}

\author{Xulin Zhao}
\affil{School of Science, Tianjin University of Technology, Tianjin, 300384, China}

\author{Dong Xu}
\affil{National Astronomical Observatories, Chinese Academy of Sciences, Beijing 100012, China}

\author{Andrea Reguitti}
\affil{Dipartimento di Fisica e Astronomia G. Galilei, Universita' di Padova, Vicolo dell'Osservaotio 3, 35122, Padova, Italy}

\author{Jujia Zhang}
\affil{Yunnan Observatories (YNAO), Chinese Academy of Sciences, Kunming 650216, China.}
\affil{Key Laboratory for the Structure and Evolution of Celestial Objects, Chinese Academy of Sciences, Kunming 650216, China.}
\affil{Center for Astronomical Mega-Science, Chinese Academy of Sciences, 20A Datun Road, Chaoyang District, Beijing, 100012, China.}

\author{Andrea Pastorello}
\affil{INAF-Osservatorio Astronomico di Padova, Vicolo dell'Osservatorio 5, 35122, Padova, Italy}

\author{Lina Tomasella}
\affil{INAF-Osservatorio Astronomico di Padova, Vicolo dell'Osservatorio 5, 35122, Padova, Italy}

\author{Paolo Ochner}
\affil{Dipartimento di Fisica e Astronomia G. Galilei, Universita' di Padova, Vicolo dell'Osservaotio 3, 35122, Padova, Italy}

\author{Leonardo Tartaglia}
\affil{Department of Astronomy, Stockholm University, SE 10691 Stockholm, Sweden}

\author{Stefano Benetti}
\affil{INAF-Osservatorio Astronomico di Padova, Vicolo dell'Osservatorio 5, 35122, Padova, Italy}

\author{Massimo Turatto}
\affil{INAF-Osservatorio Astronomico di Padova, Vicolo dell'Osservatorio 5, 35122, Padova, Italy}

\author{Avet Harutyunyan}
\affil{INAF-Fundacion Galileo Galilei, Rambla Jose Ana Fernandez Perez 7, E-38712 Brena Baja, Spain}

\author{Nancy Elias-Rosa}
\affil{INAF-Osservatorio Astronomico di Padova, Vicolo dell'Osservatorio 5, 35122, Padova, Italy}

\author{Fang Huang}
\affil{Department of Astronomy, School of Physics and Astronomy, Shanghai Jiao Tong University, Shanghai, 200240, China}
\affil{Physics Department and Tsinghua Center for Astrophysics (THCA), Tsinghua University, Beijing, 100084, China}

\author{Kaicheng Zhang}
\affil{Physics Department and Tsinghua Center for Astrophysics (THCA), Tsinghua University, Beijing, 100084, China}

\author{Juncheng Chen}
\affil{Physics Department and Tsinghua Center for Astrophysics (THCA), Tsinghua University, Beijing, 100084, China}

\author{Zhaoji Jiang}
\affil{Key Laboratory of Optical Astronomy, National Astronomical Observatories, Chinese Academy of Sciences, Beijing 100012, China}

\author{Jun Ma}
\affil{Key Laboratory of Optical Astronomy, National Astronomical Observatories, Chinese Academy of Sciences, Beijing 100012, China}
\affil{School of Astronomy and Space Science, University of Chinese Academy of Sciences}

\author{Jundan Nie}
\affil{Key Laboratory of Optical Astronomy, National Astronomical Observatories, Chinese Academy of Sciences, Beijing 100012, China}

\author{Xiyan Peng}
\affil{Key Laboratory of Optical Astronomy, National Astronomical Observatories, Chinese Academy of Sciences, Beijing 100012, China}

\author{Xu Zhou}
\affil{Key Laboratory of Optical Astronomy, National Astronomical Observatories, Chinese Academy of Sciences, Beijing 100012, China}

\author{Zhimin Zhou}
\affil{Key Laboratory of Optical Astronomy, National Astronomical Observatories, Chinese Academy of Sciences, Beijing 100012, China}

\author{Hu Zou}
\affil{Key Laboratory of Optical Astronomy, National Astronomical Observatories, Chinese Academy of Sciences, Beijing 100012, China}

\begin{abstract}
In this paper, we present extensive optical and ultraviolet (UV) observations of the type Ia supernova (SN Ia) 2013gs discovered during the Tsinghua-NAOC Transient Survey. The photometric observations in the optical show that the light curves of SN 2013gs is similar to that of normal SNe Ia, with an absolute peak magnitude of $M_{B}$ = $-$19.25 $\pm$ 0.15 mag and a post-maximum decline rate $\Delta$m$_{15}$(B) = 1.00 $ \pm $ 0.05 mag. \emph{Gehrels Swift} UVOT observations indicate that SN 2013gs shows unusually strong UV emission (especially in the $uvw1$ band) at around the maximum light (M$_{uvw1}$ $\sim$ $-$18.9 mag). The SN is characterized by relatively weak Fe~{\sc ii} {\sc iii} absorptions at $\sim$ 5000{\AA} in the early spectra and a larger expansion velocity ($v_{Si}$ $\sim$ 13,000 km s$^{-1}$ around the maximum light) than the normal-velocity SNe Ia. We discuss the relation between the $uvw1-v$ color and some observables, including Si~{\sc ii} velocity, line strength of Si~{\sc ii} $\lambda$6355, Fe~{\sc ii}/{\sc iii} lines and $\Delta m_{15}$(B). Compared to other fast-expanding SNe Ia, SN 2013gs exhibits Si and Fe absorption lines with similar strength and bluer $uvw1-v$ color. We briefly discussed the origin of the observed UV dispersion of SNe Ia.

\end{abstract}


\section{Introduction}

Type Ia supernovae (SNe Ia) have been successfully used as excellent distance indicators to measure expansion history of the universe  \citep{riess98, per99} because of their high luminosities and standardizable properties through empirical relations between their peak luminosity and light/color-curve shapes (e.g., \citealt{phi93, wxf05, guy05}). It is commonly believed that the explosion of a SN Ia arises from a carbon-oxygen white dwarf (WD) in a binary system that reaches the Chandrashekhar mass limit ($\sim$ 1.4 $M_{\odot}$; \citealt{chand57,wh12,maoz14}). However, the nature of the companion star of the exploding WD is still controversial, as it could be a main sequence (MS) star, a red-giant (RG) star, a helium star, or another WD. The former three cases correspond to the single degenerate (SD) scenario \citep{whe73}, while the last one corresponds to the double degenerate (DD) scenario \citep{iben84, web84}. Recent studies support both SD \citep{pat07, ster11, dil12} and DD scenarios \citep{schae12, edw12, sant15}.

Observations indicate that the majority of SN Ia explosions belong to the spectroscopically normal or ``Branch-normal" subclass \citep{bran93}, while the others are somewhat peculiar, including the overluminous SN 1991T-like \citep{fili97}, subluminous SN 1991bg-like \citep{fili92}, and the outstanding SN 2002cx-like \citep{liw03} subclasses. Multiple criteria have been used to subclassify such explosions. \citet{ben05} defined high-velocity-gradient (HVG) and low-velocity-gradient (LVG) subgroups, based on the differences in the velocity evolution (or gradient). \citet{bran06} used the equivalent width (EW) of Si~{\sc ii} $\lambda$6355 and Si~{\sc ii} $\lambda$5972 absorptions to classify the SNe Ia into four subgroups: core normal (CN), broad line (BL), cool (CL), and shallow silicon (SS) subclasses. The latter two sub-classes have a large overlap with peculiar SN 91bg-like and 91T-like subclasses, respectively. Using the Si~{\sc ii} velocity measured at around the maximum light, \citet{wxf09a} divided the SNe Ia into the high-velocity (HV) and the normal-velocity (NV) groups. Some subclasses from different categorizations overlap with each other. For example, HV SNe Ia have similar features as HVG and the BL ones. \citet{wxf13} compare the SN location in their host galaxy and found that HV SNe Ia are found in more luminous host environments than NV SNe Ia, suggesting that they may originate from different progenitor systems.

Ultraviolet (UV) observations play an important role in understanding diversities of SNe Ia, as the UV emission may be sensitive to the progenitor metallicity (e.g., \citealt{hof98, len00, tim03, sau08}) and/or the interaction of exploding ejecta with the companion star (i.e. \citealt{kas10}). The UV spectrum of SNe Ia, which is formed in the outer unburned C + O layer, can be attenuated by the absorption lines from the iron-peak elements (Fe~{\sc ii}, etc.).  However, there is no consensus on the metallicity effect on the UV flux. Based on the pure deflagration model W7 \citep{nom84} for SNe Ia, the UV emission increases with decreasing progenitor metallicity \citep{len00}. \citet{sau08} proposed that the reverse fluorescence scattering of photons from red to blue wavelengths can lead to stronger UV flux with increasing the iron-group elements in the outer layers. \citet{wxf12} reported that SN 2004dt showed an unusually strong UV excess. This relatively high UV flux could also be used to set constraints on the surrounding circumstellar material (CSM). On the other hand, the UV-optical colors have been used to divide the SNe Ia into several subclasses \citep{mil13}, but their scatter is larger than the values predicted by models with various metallicities \citep{brown15}. It is thus important to explore the possible origin of the intrinsic dispersion of UV luminosity in SNe Ia. Well-sampled UV observations may help us constrain the properties of progenitor system and the explosion mechanism for the different Type Ia subclasses. Moreover, multi-band light curves are also required to accurately estimate the extinction and hence the intrinsic luminosity of SNe Ia.

In this paper, optical/UV photometric and optical spectroscopic observations of SN 2013gs are presented. Observations and data reductions are described in $\S$ 2. $\S$ 3 discusses the light/color curves and the reddening estimate, while $\S$ 4 presents the spectral evolution. We discuss the relationship between the UV flux and some SNe Ia parameters in $\S$5.

\section{Observations and Data reduction}

SN 2013gs was detected on November 29.84 UT in an unfiltered image taken by the 0.6-m Schmidt telescope in the course of the THU-NAOC Transient Survey (TNTS; \citealt{zhangtm15}). The coordinates of this object are $\alpha$ = 09$^{\rm{h}}$31$^{\rm{m}}$08$^{\rm{s}}$.87, $\delta$ = +46$^{\circ}$23$\arcmin$05$\arcsec$.4 (J2000.0), located 15$\arcsec$.0 east and 2$\arcsec$.0 north of the center of an SBbc galaxy UGC 5066 (See Figure 1) with an unfiltered magnitude of about 17.3 mag. Nothing was detected in the image taken on 2013 Nov. 25 UT, with an upper detection limit of $\sim$ 19.5 mag. A spectrum taken at two days after the discovery is consistent with a young SN Ia at about 10 days before the maximum \citep{zhangjj13}. The expansion velocity measured from the Si~{\sc ii} $\lambda$6355 absorption is about 17,500 km s$^{-1}$, which is higher than that of a normal SN Ia at a similar phase.

\subsection{Photometry}

Our photometric observations of SN 2013gs began on 2013 Nov. 30 (one day after discovery) and lasted for about 4 months. The optical photometry is mainly obtained with the 0.8-m Tsinghua-NAOC reflecting Telescope (TNT\footnote{This telescope is co-operated by Tsinghua University and the National Astronomical Observatories of China (NAOC)}) located at NAOC Xinglong Observatory. A 1340 $\times$ 1300 pixel back-illuminated CCD, with a field of view (FoV) of 11.5$\arcmin$ $\times$ 11.2$\arcmin$ (pixel size $\sim$ 0.52$\arcsec$ pixel$^{-1}$) is mounted on the Cassegrain focus of the telescope \citep{huang12}. Some data were also obtained with the Yunnan Faint Object Spectrograph and Camera (YFOSC) mounted on the Li-Jiang 2.4-m telescope of Yunnan Astronomical Observatories (LJT), the 2.0-m Liverpool Telescope (LT)+IO:O and the Asiago Faint Object Spectrograph \& Camera (AFOSC) on the 1.82-m Copernico telescope\footnote{This telescope is operated by INAF Osservatorio Astronomico di Padova at Asiago, Italy.}. The UV observations were triggered immediately after the discovery, with a cadence of 2 days, using the Ultra-Violet/Optical Telescope (UVOT; \citealt{rom05}) on the \emph{Gehrels Swift} spacecraft \citep{geh04}. The UV photometry was taken from the \emph{Gehrels Swift} Optical/Ultraviolet Supernova Archive\footnote{http://swift.gsfc.nasa.gov/docs/swift/sne/swift\_sn.html} (SOUSA; \citealp{brown14}).  The reduction is based on that of \citet{brown09}, including subtraction of the host galaxy count rates and uses the revised UV zeropoints and time-dependent sensitivity from \citet{bree11}.

As SN 2013gs is located near the center of UGC 5066, the contamination of the host galaxy cannot be neglected when measuring the SN flux. Therefore, image subtraction was applied before performing photometric measurements with template images taken about one year after the explosion. The SN image is first registered to the template image, and the former is then scaled using the foreground stars to the same point spread function (PSF) as the latter. The TNT instrumental magnitudes of the SN and local standard stars are finally obtained with an ad hoc pipeline adapted for the TNT data (based on the IRAF\footnote{IRAF, the Image Reduction and Analysis Facility, is distributed by the National Optical Astronomy Observatories, which are operated by the Association of Universities for Research in Astronomy (AURA), Inc., under cooperative agreement with the National Science Foundation.} DAOPHOT package; \citealt{stet87}). The LT and AFOSC data were obtained using a python-based pipeline\footnote{http://sngroup.oapd.inaf.it/foscgui.html} \citep{cap15}.

The $UBVRI$ instrumental magnitudes of TNT and YFOSC were transformed to the standard Johnson $UBV$ \citep{john66} and Kron-Cousins $RI$ \citep{cou81} systems by observing a series of standard stars \citep{land92} on photometric nights. The $BV$ instrumental magnitudes of LT and AFOSC are calibrated with the Landolt standard stars obtained in 8 photometric nights and the $URI$ magnitudes are converted from the Sloan catalogues using the relation of \citet{cho08}. The calibrated magnitude of local standards stars are listed in Table 1. The final calibrated magnitude of SN 2013gs transformed from the instrumental magnitudes by these local standard stars are presented in Table 2. The error bars account for the uncertainties due to photon noise and photometric system calibration.

\subsection{Spectroscopy}

A total of 22 low-resolution optical spectra were obtained with different telescopes and instruments including the AFOSC on the 1.82-m Copernico telescope, Cassegrain spectrograph and BAO Faint Object Spectrograph\& Camera (BFOSC) mounted on the 2.16-m telescope at Xinglong Observatory and YFOSC on LJT  of Yunnan Astronomical Observatories. The journal of spectroscopic observations is listed in Table 3. In order to reduce the contamination from the host galaxy, the local flux next to SN 2013gs was used as background to carefully extract the SN flux which was calibrated with standard stars observed on the same night at similar air masses as the SN. The spectra were corrected for the atmospheric extinction using the extinction curves of local observatories, and telluric lines were removed using spectra of standard stars.

\section{Light Curves of SN 2013gs}

\subsection{Optical light curves from ground-based telescopes}

The $UBVRI$-band light curves of SN 2013gs are shown in Figure 2. These multi-band observations were used to derive the light-curve parameters such as peak magnitudes and the luminosity decline rate ${\Delta}m_{15}(B)$. From the polynomial fit, we find that SN 2013gs reached the $B$-band maximum on JD 2,456,640.2 $\pm$ 0.2 (2013 Dec. 13.7 UT) with $B_{max}$ = 15.53 $\pm$ 0.02 mag and the post-maximum decline rate $\Delta$m$_{15}(B)$ = 1.00 $\pm$ 0.05 mag. The light curve parameters of other bands are listed in Table 5. From the $B$- and $V$-band light curves, we derived $B_{max} - V_{max}$ = 0.17 $\pm$ 0.03 mag which suggest a non-negligible reddening in the direction of SN 2013gs.

The unfiltered light curve obtained by the 0.6-m Schmidt telescope is also reported in Figure 2 and the magnitudes are listed in Table 3. The unfiltered magnitude was calibrated by the $R$-band magnitude from the Positions and Proper Motions Star Catalogue Extended (PPMX; \citealt{ros08}). The large error bars are mainly attributed to the systematic errors between the unfiltered and $R$-band magnitudes \citep{liw03}. The unfiltered light curves reached peak of 15.01$\pm$0.20 mag on JD 2,456,641.7$\pm$1.2, about 1.5 days after $B_{max}$, while the time of first light will be estimated in section 3.4.

In Figure 3, we compared the optical light curves of SN 2013gs with other SNe Ia consistent with the SN sample used in the UV comparison, including SNe 2002bo (${\Delta}m_{15}(B)$ = 1.15; \citealt{ben04, kri04}), 2006X (${\Delta}m_{15}(B)$ = 1.17; \citealt{wxf08}), 2007gi (${\Delta}m_{15}(B)$ = 1.31; \citealt{zhangtm10}), 2009ig (${\Delta}m_{15}(B)$ = 0.89; \citealt{fol12}) and ASASSN-14lp (${\Delta}m_{15}(B)$ = 0.80; \citealt{shappee16}), 2003du (${\Delta}m_{15}(B)$ = 1.02; \citealt{stan07}), 2005cf (${\Delta}m_{15}(B)$ = 1.05; \citealt{wxf09b}). The $B$-band light curve of ASASSN-14lp is approximated from the $g$-band magnitude through the relation of \citet{cho08}. The first five SNe Ia have higher photospheric velocities and can be put into the HV subclass, while the latter two belong to the Normal subclass. The light curves of the comparison sample are  normalized to the peaks of SN 2013gs. One can see that SN 2013gs and the comparison SNe Ia have similar light curve shapes near the maximum brightness. In the early phase, SN 2013gs seems to have a rising rate similar to SN 2003du but slower than SN 2002bo, SN 2005cf and SN 2006X. This is evident in the $UBVI$ bands but not in the $R$. In the early nebular phase, the evolution of the light curve becomes more complicated. In the $B$ band, the later-time decline rate $\beta$ was estimated to be 1.39 $\pm$ 0.11 mag (100d)$^{-1}$, which is slightly smaller than those of SN 2003du (1.65 $\pm$ 0.02 mag (100d)$^{-1}$) and SN 2005cf (1.62$\pm$0.07) mag (100d)$^{-1}$ as shown in Figure 3b. The decay rate in other bands are also calculated and listed in Table 5. Although the $R$- and $I$-band light curves of SN 2013gs are similar to the comparison SNe Ia, its second bump/shoulder feature (at around t $\approx$ 25 days) is weaker than the comparison SNe Ia (in particular the HV SNe Ia).

\subsection{UV Light Curves from \emph{Gehrels Swift} UVOT}

The UV and optical light curves of SN 2013gs obtained by \emph{Gehrels Swift} UVOT are also shown in Figure 2 and listed in Table 4. SN 2013gs was not detected in the $uvm2$ band (centered at $\sim$ 2200 {\AA}) presumably due to the large reddening bump near 2200 {\AA} \citep{car89}. In Figure 4, we compared the UV ($uvw1$ and $uvw2$) light curves of SN 2013gs with SN 2005cf (${\Delta}m_{15}(B)$ = 1.05; \citealt{wxf09b,past07}), SN 2006X (${\Delta}m_{15}(B)$ = 1.17; \citealt{brown09}), SN 2009ig (${\Delta}m_{15}(B)$ = 0.89; \citealt{fol12}), SN 2011fe (${\Delta}m_{15}(B)$ = 1.18; \citealt{brown12b, zhangkc16}), SN 2012fr (${\Delta}m_{15}(B)$ = 0.80; \citealt{zhangjj14,contreras18}) and ASASSN-14lp (${\Delta}m_{15}(B)$ = 0.80; \citealt{shappee16}). The $uvw1$- and $uvw2$-band light curves of SN 2013gs span from t = $-$11.5 days to t = $+$6.5 days relative to the $B$-band maximum. The UV light curves of the comparison SNe Ia have been normalized to their $B$-band maxima and shifted to match the UV peaks of SN 2013gs. The $uvw1$-band light curve of SN 2013gs shows close resemblance to SN 2009ig around the maximum light. In comparison, SN 2006X shows a slower post-maximum decline rate while SN 2011fe declines more rapidly than other SNe Ia. The $uvw2$-band light curve of SN 2013gs is similar to that of SN 2009ig and SN 2011fe, though it has larger photometric errors. The UV light curves of SN 2013gs are similar to those of SN 2009ig.

 \subsection{The Reddening and Color curves}

The Galactic reddening toward SN 2013gs is $A_{V}^{Gal}$ = 0.063 mag from the dust map of \citet{sch98} and 0.053 mag from \citet{sf11}. We adopted the average value of 0.058 mag, corresponding to $E(B-V)$ = 0.019 mag using the standard extinction coefficient $R_{V}$ = 3.1. The reddening from the host galaxy can be estimated through the relation between ${\Delta}m_{15}(B)$ and $B_{max} - V_{max}$ (see \citealt{phi99, wxf09b}). The comparison gives a host reddening of $E(B-V)$ = 0.21 mag for SN 2013gs. With the $\Delta$$C_{12}$ method \citep{wxf05}, we estimated $E(B-V)$ = 0.22 mag. The average value of $E(B-V)_{host}$ = 0.22 $\pm$ 0.05 mag is adopted for this paper. The interstellar Na {\sc i} D doublet lines from host galaxy are not discerned clearly in the low-resolution spectra of SN 2013gs, hence cannot be used to estimate independently the host galaxy reddening. The extinction at NUV band varied slightly with different $R_{V}$. SN 2013gs is not a very high-velocity Ia like 2006X that we prefer to use $R_{V}$ = 3.1 to estimate the host galaxy extinction.

Figure 5 presents the reddening-corrected color curves of SN 2013gs. The color curves of SN 2013gs are similar to those of the comparison SNe Ia before and around the maximum light. As shown in Figure 5a and 5b, SN 2013gs reached its bluest $U - B$ and $B - V$ colors at about 7 days before the $B$-band maximum, while they reach the reddest values at about 3-4 weeks after the maximum light. We also note that SN 2002bo, SN 2006X, and SN 2007gi reach their red peaks after maximum about one week earlier than normal SNe Ia (i.e., SN 2003du and SN 2005cf), suggestive of possible correlation of this difference with ejecta velocity. After t $\sim$ 1 months, the color curves of those SNe Ia are characterized by more heterogeneous evolution. In particular, SN 2013gs is bluer than the comparison SNe Ia in $U - B$ and $B - V$, while it is redder in the $V - R$ and $V - I$ colors (Figure 5c and 5d). Another interesting feature about SN 2013gs is that its $V-R$ color seems to be bluer than the other SNe Ia at very early phases (i.e., at t $<$ -10 days).

The UV color curves of SN 2013gs and the comparison sample are shown in Figure 6. The heterogeneity of the $UV - V$ colors is larger than that of the optical colors, due to the significant change of the photospheric temperature during the explosion and/or a larger line blanketing in the UV domain. The $uvw1 - V$ color curve has a sort of a ``V" shape, with the bluest color reached at about 5 days before the $B$-band maximum. Such an "V"-shape feature is also mentioned by \citet{mil10}, and is similarly seen in the $U - B$ and $B - V$ color curves. We note that SN 2013gs has one of the bluest $uvw1 - V$ before the maximum light, similar to SN 2011fe. SN 2011fe also shows very strong UV emission \citep{zhangkc16} and belongs to the ``NUV-blue" group of SNe Ia defined by \citet{mil13}. In $uvw1 - V$, SN 2013gs is bluer than SN 2009ig by $\sim$ 0.2 - 0.4 mag, though these two SNe Ia have similar shapes in their UV light curves. SN 2013gs has the reddest $uvw2 - V$ color among all SNe Ia of our sample for the earliest data point. This deviation may be due to the large uncertainties associated with the red-leak corrections \citep{brown10, brown15}. Since SN 2013gs can be associated to both the HV and NUV-blue subclasses of SNe Ia, it apparently becomes an exception for the tendency that NUV-blue events belong to a subset of the larger low-velocity gradient group \citep{mil13}.

\subsection{The rise time}
The explosion time of SNe Ia can be inferred from the very early rise of its light curves. Some properties of progenitor systems, such as the radius of exploding star \citep{piro13} or binary interaction \citep{kas10}, can also be constrained by the early-time evolution of the light curves. The first detection of SN 2013gs was made at t = $-$13.86 days from the $B$-band maximum light and an upper limit to $\sim$ 19.5 mag was found at about t = $-$17.80 days before the $B$-band maximum. Assuming that SNe Ia expand like an homogenous fireballs at very early times, the rise time $t_{r}$, depending primarily on the opacity, could be calculated by the $t^{2}$ model ($L$ $\propto$ $t^{2}$; \citealt{arn82, riess99}). Using the data from the first detection to about 8 days before the $B$-band maximum and a $t^{2}$ model, we constrain the best-fit bolometric rise time as 18.72 $\pm$ 0.18 days. This rise time is larger than those of SN 2009ig (17.13 $\pm$ 0.07 days; \citealt{fol12}), SN 2011fe (17.64 $\pm$ 0.01 days; \citealt{zhangkc16}), SN 2013dy ($\sim$17.8 days; \citealt{zheng13}), ASASSN-14lp (16.94 $\pm$ 0.10 days; \citealt{shappee16}), the average Sloan Digital Sky Survey (SDSS) SN Ia (17.38 $\pm$ 0.17 days in $B$ band; \citealt{hay10}) and the average low-redshift SNe Ia (17.44 $\pm$ 0.39 days; \citealt{stro07}). Considering a general form of the fireball model, the index of power law can be a free parameter ($L$ $\propto$ $t^{n}$). For the unfiltered light curve as shown in Figure 2, we obtain a best-fit rise time of 16.89 $\pm$ 0.88 days and the corresponding index is 1.47 $\pm$ 0.26. This smaller index is consistent with the value given by \citet{piro13} with 1.5. For the model fit with free index, we obtain a rise time of 16.28 $\pm$ 1.25 in U, 15.81 $\pm$ 0.91 in B, 17.69 $\pm$ 1.36 in V, 17.21 $\pm$ 3.25 in R and 18.76 $\pm$ 3.03 days in I, respectively. However, the first-light time derived from the above analysis may have great uncertainty due to the lack of early time photometry.

\subsection{The absolute magnitude and quasi-bolometric light curve of SN 2013gs}

The host galaxy of SN 2013gs has a distance modulus of $m - M$ = 34.29 $\pm$ 0.15 mag \citep{mou00}. Adopting the reddening discussed above, the peak absolute magnitudes of SN 2013gs are $M_{U}$ = $-$19.62 $\pm$ 0.15 mag, $M_{B}$ = $-$19.29 $\pm$ 0.15 mag, $M_{V}$ = $-$19.24 $\pm$ 0.15 mag, $M_{R}$ = $-$19.17 $\pm$ 0.15 mag ,and $M_{I}$ = $-$18.90 $\pm$ 0.15 mag. They are similar to those of typical SNe Ia (e.g. $M_{V}$ = $-$19.27 mag from \citet{wxf06}).  The UV-band peak absolute magnitude are also calculated as $M_{uvw1}$ = $-$18.86 $\pm$ 0.18 mag and $M_{uvw2}$ = $-$17.87 $\pm$ 0.22 mag.

We build the quasi-bolometric light curve of SN 2013gs based on the UV and optical photometry to estimate the peak bolometric luminosity and nickel mass produced in the explosion. The bolometric light curves of SN 2013gs are mostly determined by the optical photometry. The early UV contribution is derived from the \emph{Gehrels Swift} UVOT observations. The ratio between UV ($\sim$ 1500 - 3000 {\AA}) and optical emission ($\sim$ 3000 - 9000 {\AA}) is about 0.15 in the rising phase and declined to 0.12 at around the $B$-band maximum light larger than those of SN 2005cf and SN 2011fe ($\sim$ 10\%; \citealt{wxf09b, zhangkc16}) at similar phases. We did not observe SN 2013gs in the near-infrared (NIR) bands, and a contribution of $\sim$ 5\% (derived from SN 2005cf at around the maximum light) is thus applied to the NIR correction of bolometric luminosity. For SN 2013gs, the peak bolometric luminosity is estimated to be 1.82 ($\pm$0.16) $\times$ 10$^{43}$ erg s$^{-1}$, which is comparable to that of other SNe Ia. The error includes uncertainties in distance modulus, photometry, reddening correction, and flux correction beyond the optical window.

Using the Arnett law \citep{arn82}, we can estimate the mass of radioactive $^{56}$Ni from the maximum-light luminosity using the following relation:
\begin{equation}
L_{\rm{max}} = (6.45 \times e^{\frac{-t_{r}}{8.8}}+1.45 \times e^{\frac{-t_{r}}{111.3}}) ~M_{\rm{Ni}} \times 10^{43} ~\rm{erg} ~ \rm{s^{-1}},
\end{equation}
where $t_{r}$ is the rise time of the bolometric light curve, and $M_{\rm{Ni}}$ is the mass of $^{56}$Ni in unit of solar mass \citep{stri05}. For SN 2013gs, $t_{r}$ is taken as 16.89 $\pm$ 0.88 days from the analysis in the above subsection. Inserting the parameters of $t_{r}$ and L$_{max}$ into the above equation, we derive a nickel mass of as 0.83 $\pm$ 0.10 $M_{\odot}$. This value is within the range of normal SNe Ia.

\section{Optical spectra}

A total of 22 optical spectra were taken for SN 2013gs with different telescopes, spanning from t = $-$12 days to +85 days relative to the $B-$band maximum. The journal of spectroscopic observations is \text{shown} in Table 6. The complete spectral evolution is presented in Figure 7, where the spectra at the maximum brightness are characterized by broad, blueshifted absorption lines of Ca {\sc ii} H \& K, Si~{\sc ii} $\lambda$6355 and the Ca NIR triplet. The detailed spectral evolution is discussed in the following subsections.

\subsection{Temporal Evolution of the Spectra}

In Figure 8, we compare the spectra of SN 2013gs with well-observed SNe Ia with similar ${\Delta}m_{15}(B)$ at representative epochs (t $\sim$ $-$10, 0, +1 week and +1 month since the $B$ maximum). The sample includes SN 2002bo \citep{ben04}, SN 2003du \citep{stan07},  SN 2004dt \citep{alta07, silver12}, SN 2005cf \citep{wxf09b} and SN 2009ig \citep{fol12}. All spectra are corrected for the reddening of Milky Way and host galaxies.

The earliest spectrum of SN 2013gs taken at t $\approx$ $-$12 days (Figure 8a) shows singly ionized lines of intermediate-mass elements (IMEs) like Si, S, Mg and Ca. The O~I $\lambda$7773 is not detected in the spectrum of SN 2013gs at any epoch. Compared to other HV SNe Ia, the absorption lines of Si~{\sc ii} $\lambda$6355 and W-shaped S {\sc ii} features are relatively weak in SN 2013gs. The absorption at about 5800~{\AA} due to Si~{\sc ii} $\lambda$5972 is barely visible in SN 2013gs, while it is prominent in SN 2002bo at comparable phases. The Fe~{\sc ii}/{\sc iii} absorption ($\sim$ 5000{\AA}) is shallower than some HV SNe Ia such as SN 2002bo. SN 2013gs show similar Ca~NIR triplet with SN 2004dt, but the latter has a clear O~{\sc i} absorption, an important spectral diagnostic for constraining the explosion mechanism of SNe Ia \citep{zhao16}. At around the maximum brightness (Figure 8b), SN 2013gs is quite similar to SN 2002bo, but its spectral features are weaker, especially for the Ca {\sc ii} NIR triplet. We measure \textsl{R}(Si~{\sc ii}) \citep{nug95}, the line ratio of Si~{\sc ii} $\lambda$5972 and Si~{\sc ii} $\lambda$6355, to be 0.08 $\pm$ 0.02 for SN 2013gs. This smaller value of \textsl{R}(Si~{\sc ii}) suggests that SN 2013gs has a higher photospheric temperature. It is expected that the spectral energy distribution (SED) of SN shifts towards shorter wavelengths with higher photospheric temperatures. The higher temperature can also keep the iron-group elements at higher ionization states, which finally results in less degree of line blanketing in the UV region \citep{sau08}. This could produce stronger UV emission for SN 2013gs. By one week after maximum light (Figure 8c), the W-shaped profile of S {\sc ii} line almost disappeared in SN 2013gs and other HV SNe Ia, whilst this feature is still visible in SN 2005cf. At t $\sim$ 1 month (Figure 8d), the spectral profiles of all SNe Ia become quite similar to each other, although the absorptions of Ca~{\sc ii} NIR triplet still shows some diversity in strength.

\subsection{Photospheric Expansion Velocity}

Figure 9 shows the expansion velocities of SN 2013gs inferred from the Si~{\sc ii} 6355 and Ca~{\sc ii} NIR triplet absorptions, along with those obtained for SNe 2002bo, 2003du, 2005cf, 2006X, 2007gi, 2009ig and 2011fe \citep{zhao15, silver15}. The Si~{\sc ii} velocity evolution of SN 2013gs is very close to that of SNe 2002bo and 2009ig, lying between the fast-expanding objects like SN 2006X and the normal SNe Ia like SNe 2005cf and 2011fe. The profiles of Si~{\sc ii} $\lambda$6355 from t = $-$12 days to t = +5 days are symmetric and can be fitted well with a single-Gaussian function. No high-velocity features are observed in the spectral sequence. Assuming that the expansion velocity of the photosphere decreases in an exponential way, we estimate the Si~{\sc ii} velocity of SN 2013gs as v$_{si}$ = 12,800 $\pm$ 400 km s$^{-1}$ at around the maximum light, with a velocity gradient $\dot{v}$ = 101 $\pm$ 10 km  s$^{-1}$. These values are noticeably larger than those of normal SNe Ia, putting SN 2013gs into the HV or HVG subclasses of SNe Ia according to the classification schemes proposed by \citet{wxf09b} and \citet{ben05}. The velocities inferred from the S~{\sc ii}, and Ca~{\sc ii} are higher than Normal SNe Ia by 1,000 km s$^{-1}$, 2,000 km s$^{-1}$, respectively. At t~=~$-$7 days, the absorption profiles of Ca {\sc ii} H $\&$ K are asymmetric, probably as a consequence of contamination from Si~{\sc ii} $\lambda$3860. It is hard to distinguish the HV component and measure its strength. The absorption line almost keeps a Gaussian profile and varies slowly since t = $-$5 days. The Ca~{\sc ii} NIR triplet display a triangle-shaped profile in the spectra of SN 2013gs before t~=~$-$5 days. The substructures of Ca {\sc ii} $\lambda$8498, $\lambda$8542, and $\lambda$8662 absorptions are blended. In analogy with Si~{\sc ii} $\lambda$6355 and Ca H $\&$ K, the profile of Ca {\sc ii} NIR triplet evolves to become symmetric after t = $-$5 days.

\section{Dispersion of the Observed UV flux}

The UV emission of SNe Ia produced in deeper layers are blanketed by the bound-bound transitions from iron-group elements. Thus, most of the UV emission of SNe Ia originates from the outer layers of the ejecta. One of the remarkable characteristics in the spectra of SNe Ia is that the feature of Si~{\sc ii} $\lambda$6355 is theoretically proposed to vary with the metallicity in the C~+~O layer. From the synthetic spectra of W7 model, the increasing metallicity could lead to a blueward shift of the Si~{\sc ii} absorption and a decrease in the UV pseudo-continuum \citep{len00, sau08, walk12}. We calculate the expansion velocity and pseudo-equivalent width (pEW) of SN 2013gs at its maximum, and compare with other SNe Ia with different UV-optical color. The UV-optical peak color of comparison SNe Ia mainly come from \citet{brown10, brown17}. These sample, together with SNe 2004dt\footnote{The $uvw1 - v$ colors of SN 2004dt are converted from the HST UV photometry.}\citep{wxf12}, 2008hv \citep{brown12a}, 2009ig \citep{fol12}, 2011fe \citep{brown12b}, 2012fr \citep{zhangjj14, contreras18}, ASASSN-14lp \citep{shappee16} from the literatures, and 2013gs presented in this paper, are all corrected for their total reddening.

Figure 10 presents the relations between $uvw1 - v$, $b - v$ colors (panel a), $\Delta$m$_{15}$(B) (panel b), Si {\sc ii} velocity (panel c), pEW of Si~{\sc ii} $\lambda$6355 (panel d) and Fe~{\sc ii}/{\sc iii} ($\sim$ 5000{\AA}) absorptions (panel e) for these SNe Ia. The colors, velocities and pEWs are measured around $B$-band peak. The velocity of Si~{\sc ii} $\lambda$6355 ($v_{Si}$ = 12000 km s$^{-1}$) allows us to separate the sample into the HV and NV subclasses. The $uvw1 - v$ and $b - v$ color plots are shown in Figure 10a. The $(b - v)_{BPEAK}$ of NUV-blue events is bluer than the NUV-red group as mentioned by \citet{mil13}. In Figure 10b, we plot the $uvw1 - v$ color versus the $B$-band post-maximum decline rate $\Delta$m$_{15}$(B) for SN 2013gs and some other SNe Ia with UV observations. The HV SNe Ia show a trend that faster decliners have weaker UV fluxes. On the contrary, the $uvw1 - v$ color of NV SNe Ia has large dispersion and they do not show correlation between $uvw1 - v$ and $\Delta$m$_{15}$(B). \citet{wxf13} suggested that the progenitors of HV SNe Ia are likely associated with younger and metal-rich stellar environments. Thus the UV emission of SNe Ia may be more sensitive to the metallicity of its progenitors. The different behaviors seen in the observed UV flux of SNe Ia may be related to the difference in their progenitors and explosion models. A recent study by \citet{brown17} suggests SNe Ia with higher ejecta velocities tend to have redder $uvw1 - v$ color. However, SN 2004dt, ASASSN-14lp and SN 2013gs in our sample do not seem to follow this trend, as shown in Figure 10c. These outliers provide evidence that the observed UV flux of SNe Ia may be affected by multiple parameters, including metallicity, asymmetric explosions and circumstellar interaction \citep{brown15}. For instance, the shocked ejecta and/or CSM may be compressed into a cool dense shell during the first week after the explosion, which can absorb the X-rays produced by the shock and then reradiate in the UV domain (\citealt{fran84}, etc.). However, the typical intermediate-width Balmer or helium lines produced by the interaction are not detected in the spectra of these SNe Ia. The asymmetric explosion of SNe Ia will create the inhomogeneities in the ejecta density and structure. \citet{kp07} used the detonation from failed deflagration (DFD) model to create synthetic spectra observed from different viewing angles. The aspherical distribution of iron-group elements after explosion may also lead to the stronger flux variation at the ultraviolet than the optical bands. In Figure 10d and 10e, One may see that HV SNe Ia tend to have large pEW of Si {\sc ii} and Fe {\sc ii}/{\sc iii} absorptions relative to the NV ones, as also indicated by previous studies. The trend that lower metallicity results in smaller line velocities was established by \citet{len00} based on the W7 model. Meanwhile, \citet{len00} also suggested that the decrease in metallicity leads to increasing UV flux, but with no effects in the optical continuum \citep{sau08}.

\section{Summary}

In this paper, extensive optical and ultraviolet observations are presented for the Type Ia SN 2013gs discovered in UGC 05066 (z $\sim$ 0.0169) by the Tsinghua-NAOC Transient Survey. The optical photometry started from t $\sim$ $-$14 days before the $B$-band maximum and extended to t $\sim$ $+$110 days. With the unfiltered and multi-band light curves, we estimate the peak B-band magnitude as $B_{max}$ = 15.53 $\pm$ 0.02 mag, ${\Delta}m_{15}(B)$ = 1.00 $\pm$ 0.05 mag, $B_{max} - V_{max}$ color as 0.15 $\pm$ 0.03 mag, and a rise time of 16.89 $\pm$ 0.88 days. The host-galaxy reddening is $E( B - V )$ = 0.22 $\pm$ 0.05 mag through several empirical methods, from which we estimate the absolute peak magnitude as $M_{B}$ = $-$19.25 $\pm$ 0.15 mag. The overall evolution of the light curve of SN 2013gs is similar to that of SN 2003du.

The UV light curves of SN 2013gs do not match well other SNe Ia, especially at later phase. The reddening-corrected UV colors of SN 2013gs show an excess relative to other HV SNe Ia. This phenomenon is also noticed in another HV SN Ia -- SN 2004dt. We estimate the ejected $^{56}$Ni mass of $\sim$ 0.8 $M_{\odot}$ using the quasi-bolometry light curve of SN 2013gs. The spectroscopic behavior of SN 2013gs is consistent with that of typical HV SNe Ia. The Si~{\sc ii} absorption velocity $v_{\rm{exp}}$ is found to be 12,800 $\pm$ 400 km s$^{-1}$, which indicates that SN 2013gs can be put into the category of HV subclass. The spectral and velocity evolution of SN 2013gs is similar to that of HV SNe Ia like SN 2002bo. The pEW of Si~{\sc ii} $\lambda$6355 and Fe {\sc ii}/{\sc iii} ($\sim$ 5000{\AA}) measured for SN 2013gs is comparable to that obtained for other HV SNe near the maximum light, but it has bluer $uvw1 - V$ colors. Such a strong UV emission is also seen in another two HV Ia SN 2004dt and ASASSN-14lp. This further complicates our understanding of the discrepancy in the UV emission. For the HV SNe Ia, those with slower decline rate ($\Delta$m$_{15}$(B)) tend to have stronger UV emission, while this tendency does not hold for the sample of NV SNe Ia.

\acknowledgments This work is supported by the National Natural Science Foundation of China (NSFC grants 11178003, 11325313, and 11633002), Strategic Priority Research Program of the Chinese Academy of Sciences, Grant No. XDB023000000. This work is partially supported by the National Natural Science Foundation of China (NSFC grants 11603034, 11433005 and 11673027) and the External Cooperation Program of Chinese Academy of Sciences (Grant No. 114A11KYSB20160057). Funding for the LJT has been provided by the Chinese Academy of Sciences (CAS) and the People's Government of Yunnan Province. The LJT is jointly operated and administrated by Yunnan Observatories and Center for Astronomical Mega-Science, CAS. J.-J. Zhang is supported by NSFC (grants 11403096 and 11773067), the Youth Innovation Promotion Association of the CAS, the Western Light Youth Project, and the Key Research Program of the CAS (Grant NO. KJZD-EW-M06). We acknowledge the support of the staff of the Xinglong 2.16m/80cm/Schmidt telescopes. This work was partially supported by the Open Project Program of the Key Laboratory of Optical Astronomy, National Astronomical Observatories, Chinese Academy of Sciences. AP, LT, SB, PO and MT are partially supported by the PRIN-INAF 2017 towards the SKA and CTA era: discovery, localisation and physics of transient sources (PI M. Giroletti).

\clearpage

\begin{figure}[ht]
\centering
\includegraphics[angle=0,width=120mm]{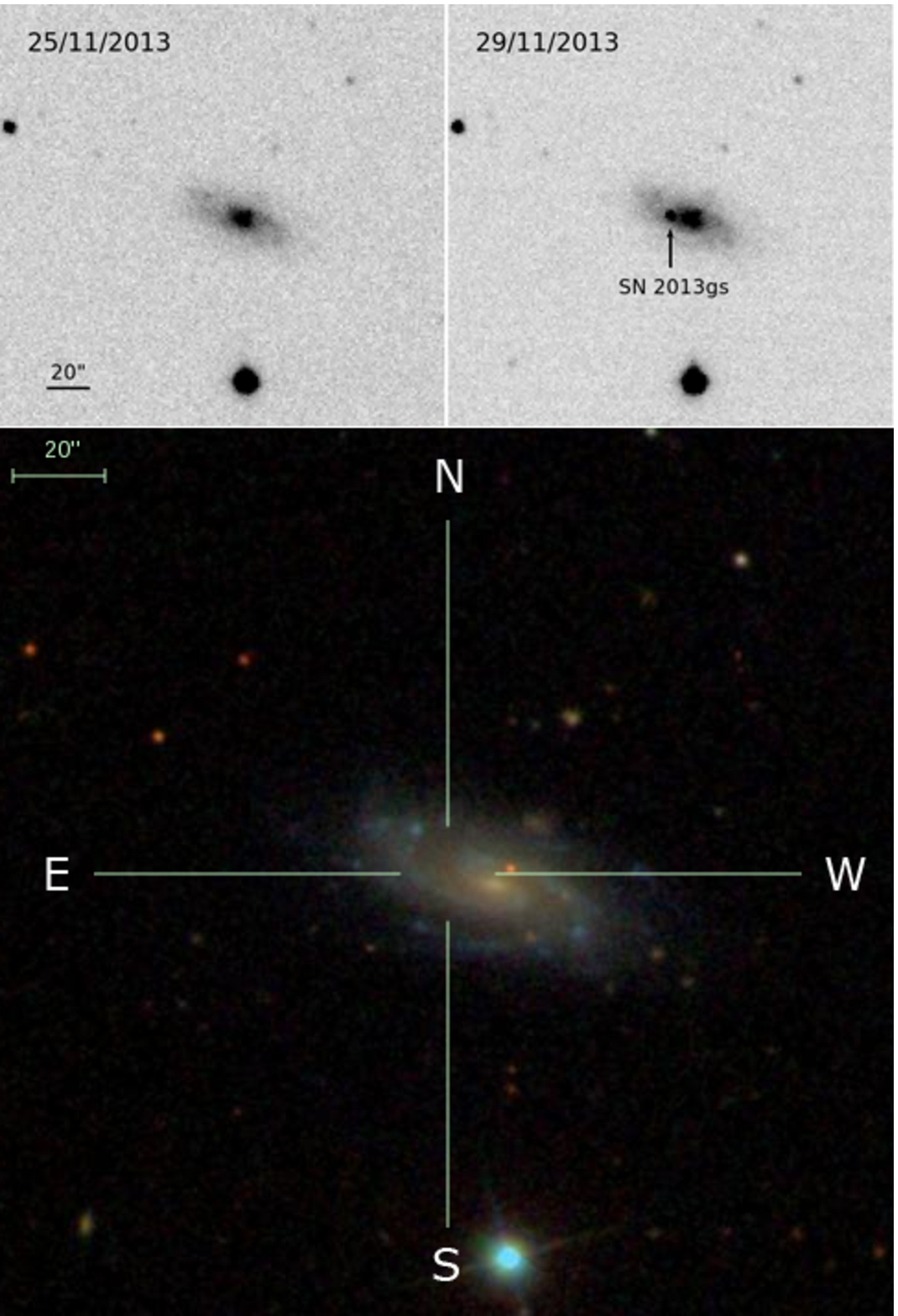}
\caption{Top left: the unfiltered image without SN 2013gs taken by TNTS on Nov. 24th. Top right: the unfiltered image with SN (marked by the arrow) taken by TNTS on Nov. 29th. Bottom: SDSS color image of the host galaxy of SN 2013gs and the position of SN is marked by the crosshair. North is up and east is to the left (A color version of this Figure is available in the online journal.)}
\end{figure}

\begin{figure}[ht]
\centering
\includegraphics[angle=0,width=150mm]{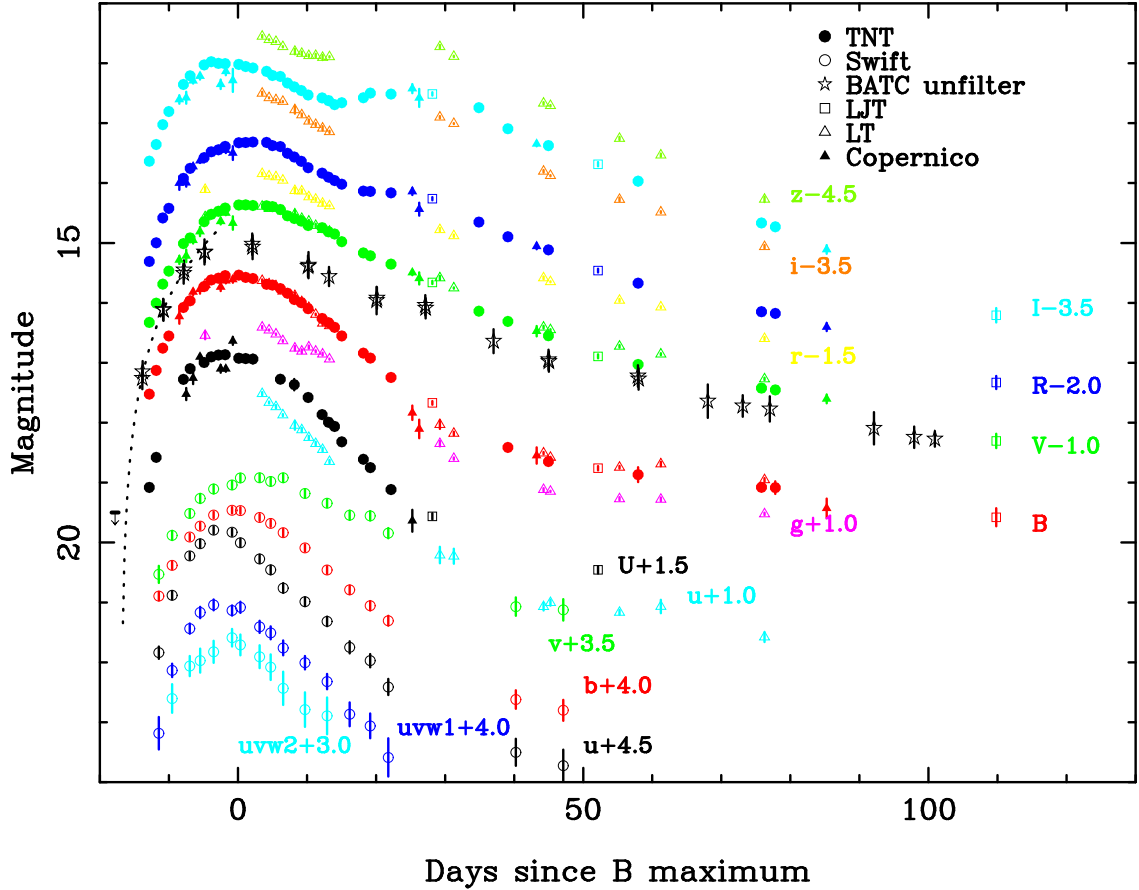}
\caption{UV and optical light curves of SN 2013gs. The two UV ($uvw1$ and $uvw2$) and three optical ($u$, $b$ and $v$) light curves were obtained by \emph{Gehrels Swift} UVOT (open circles). The filled circles are the $UBVRI$ light curves obtained by TNT. The data coming from LJT 2.4-m telescope (open squares), 2.0-m Liverpool telescope (open triangles) and 1.82-m Copernico Telescope (filled triangles) are also plotted in the figure. The open pentagons shows the unfilter light curves of the 0.6-m Schmidt telescope. The light curves are shifted vertically for clarify. The dashed black line is the $t^{1.47}$ fit for unfilter data before -8 days relative to the $B$-band maximum.}
\end{figure}

\begin{figure}[ht]
\centering
\includegraphics[angle=0,width=120mm]{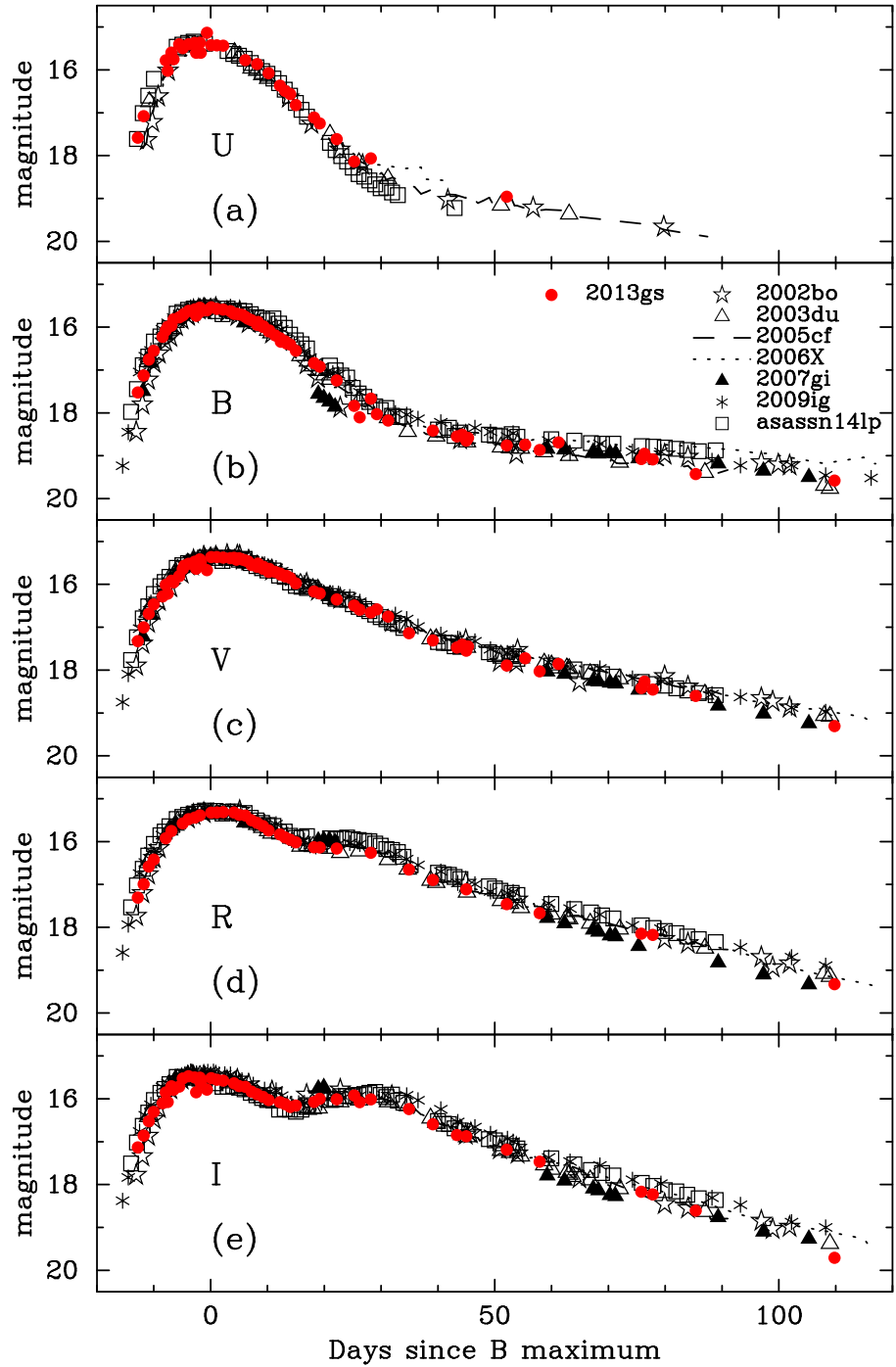}
\caption{$UBVRI$ light curves of SN 2013gs, and the comparison SNe 2002bo, 2003du, 2005cf, 2006X, 2007gi, 2009ig and ASASSN-14lp (see text for the references). All light curves are shifted in magnitude to match the peak of SN 2013gs.}
\end{figure}

\begin{figure}[ht]
\centering
\includegraphics[angle=0,width=120mm]{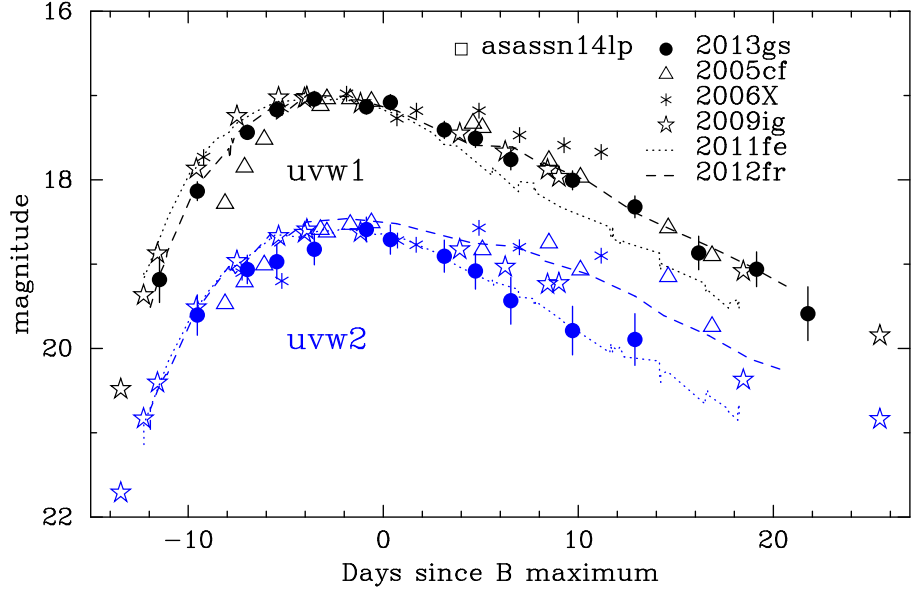}
\caption{UV light curves of SN 2013gs observed with \emph{Gehrels Swift} UVOT ($uvw1$ and $uvw2$), compare with SNe 2005cf, 2006X, 2009ig, 2011fe and 2012fr (see text for the references). All light curves are shifted in magnitude to match the peak of SN 2013gs.}
\end{figure}

\begin{figure}[ht]
\includegraphics[angle=0,width=150mm]{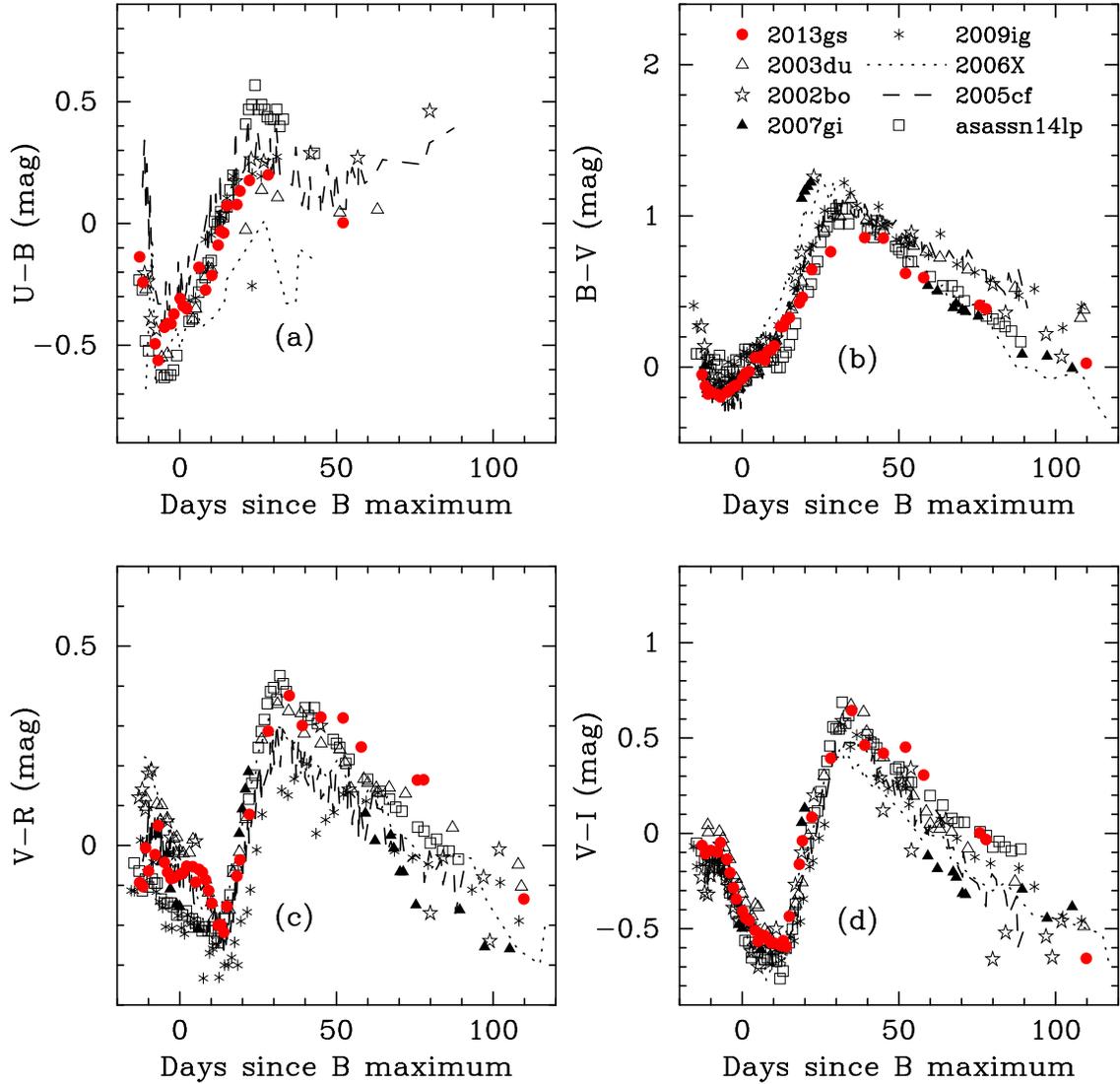}
\caption{$U - B$, $B - V$, $V - R$, and $V - I$ color curves of SN 2013gs compared with those of SNe 2002bo, 2003du, 2005cf, 2006X, 2007gi, 2009ig and ASASSN-14lp. All SNe have been dereddened.}
\end{figure}

\begin{figure}[ht]
\centering
\includegraphics[angle=0,width=120mm]{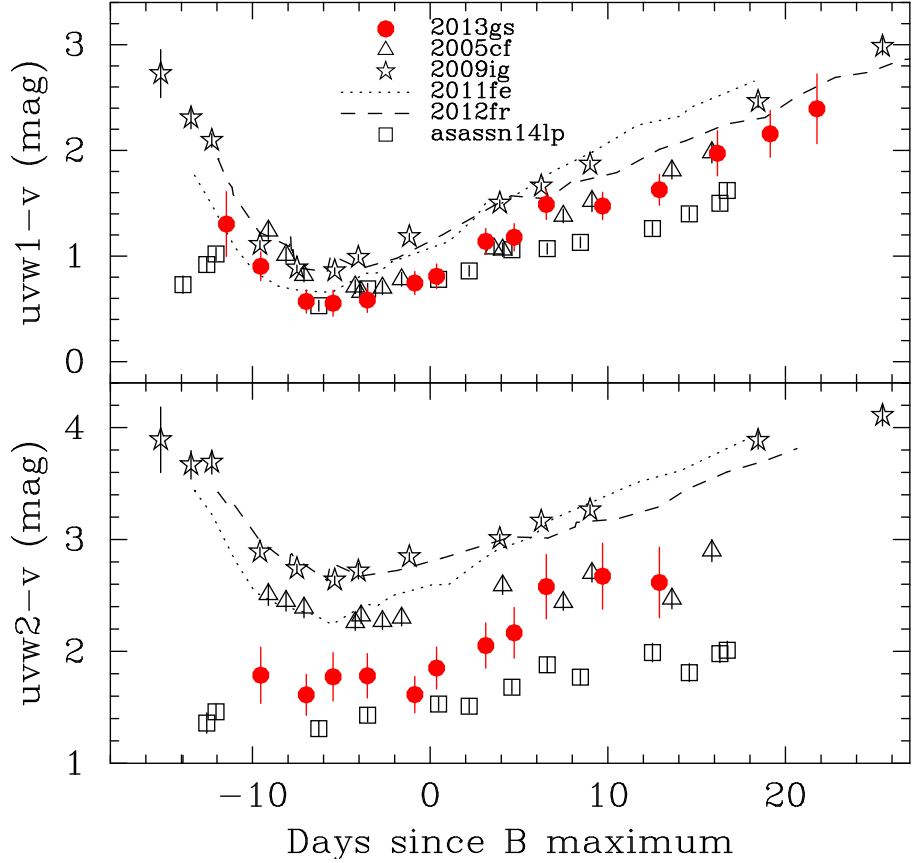}
\caption{$uvw1 - v$ and $uvw2 - v$ colors of SN 2013gs compared with those of SNe 2005cf, 2006X, 2009ig, 2011fe, 2012fr and ASASSN-14lp (see text for the references). All SNe have been dereddened and after red-tail correction.}
\end{figure}

\begin{figure}[ht]
\centering
\includegraphics[angle=0,width=120mm]{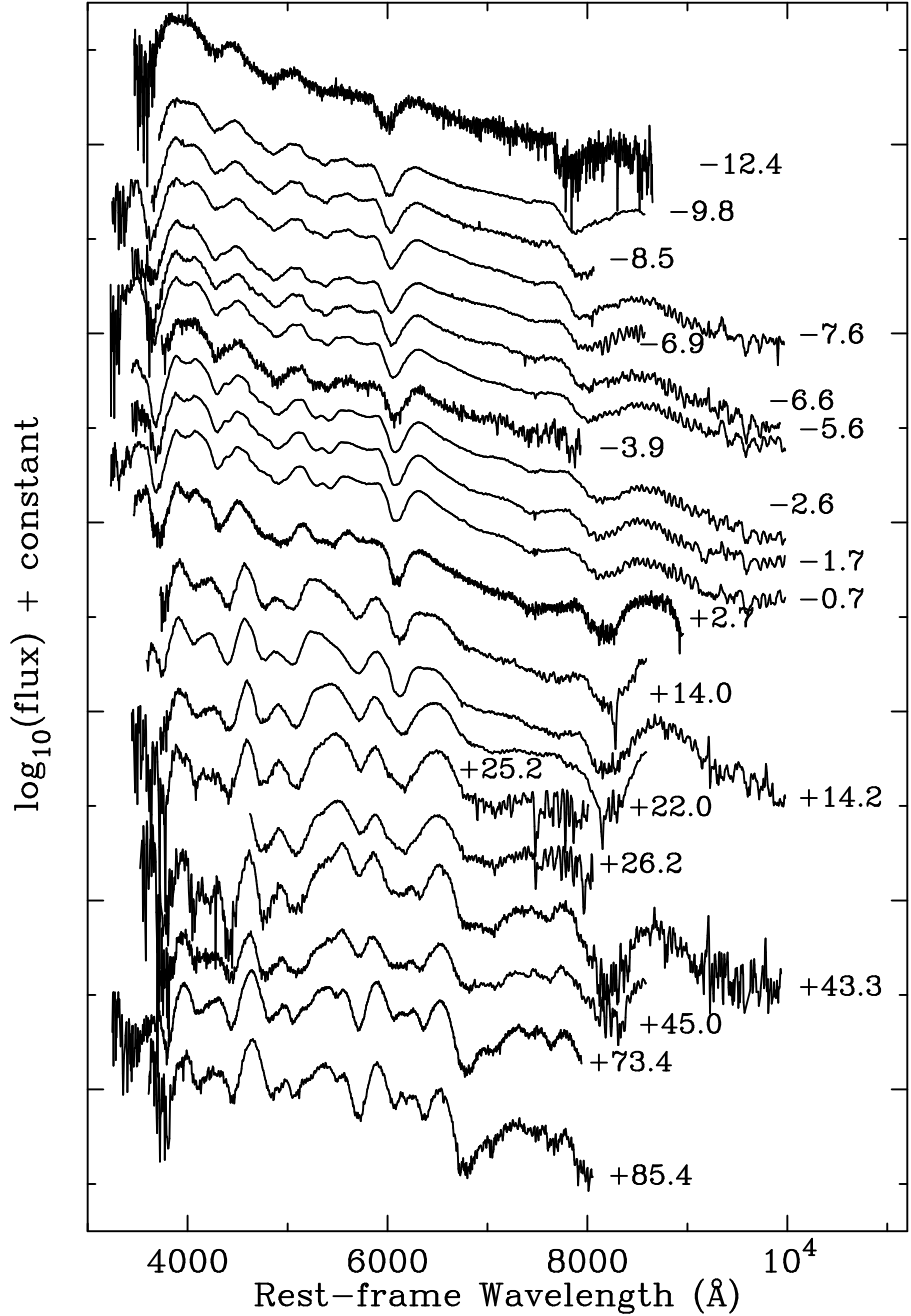}
\caption{Optical spectral evolution of SN 2013gs. All the spectra have been corrected for the redshift of UGC 05066 ($v_{\rm{hel}}$ = 5063 km s$^{-1}$) but only the reddening of Milky Way. The spectra are arbitrarily shifted in vertical direction for the clarity.}
\end{figure}

\begin{figure}[ht]
\centering
\includegraphics[angle=0,width=150mm]{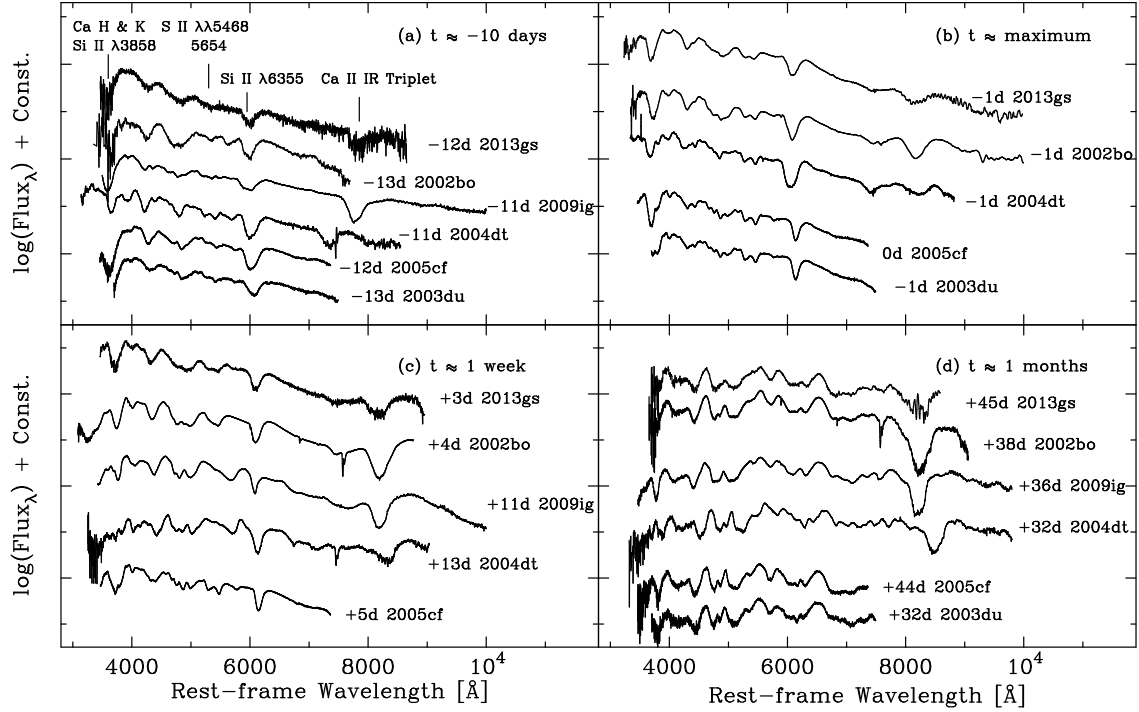}
\caption{Spectra of SN 2013gs at four different epochs (a-d, t $\approx$ $-$12 days, $-$1 days, +3 days and 1 months from $B$-band maximum), shown along with spectra of other SNe 2002bo, 2003du, 2004dt, 2005cf and 2009ig at similar epochs (see the text for the references). All spectra have been corrected for the redshift of host galaxy and shifted in the vertical direction for better display.}
\end{figure}

\begin{figure}[ht]
\centering
\includegraphics[angle=0,width=150mm]{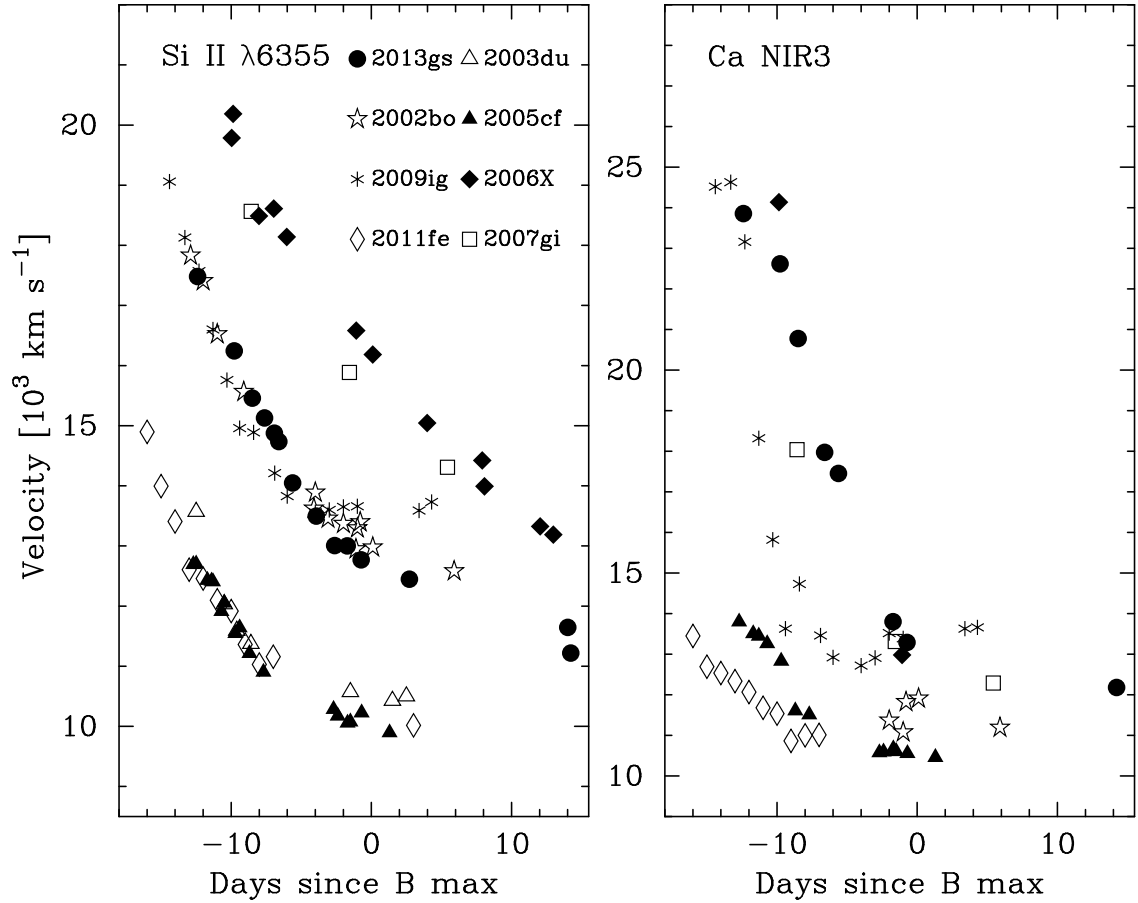}
\caption{Evolution of the expansion velocity of SN 2013gs measured from the minimum of Si~{\sc ii} $\lambda$ 6355 (left panel) and Ca NIR triplet (right panel) absorptions , compared with those of SNe 2002bo, 2003du, 2005cf, 2006X, 2007gi, 2009ig and 2011fe (see text for the references).}
\end{figure}

\begin{figure}[ht]
\centering
\includegraphics[angle=0,width=120mm]{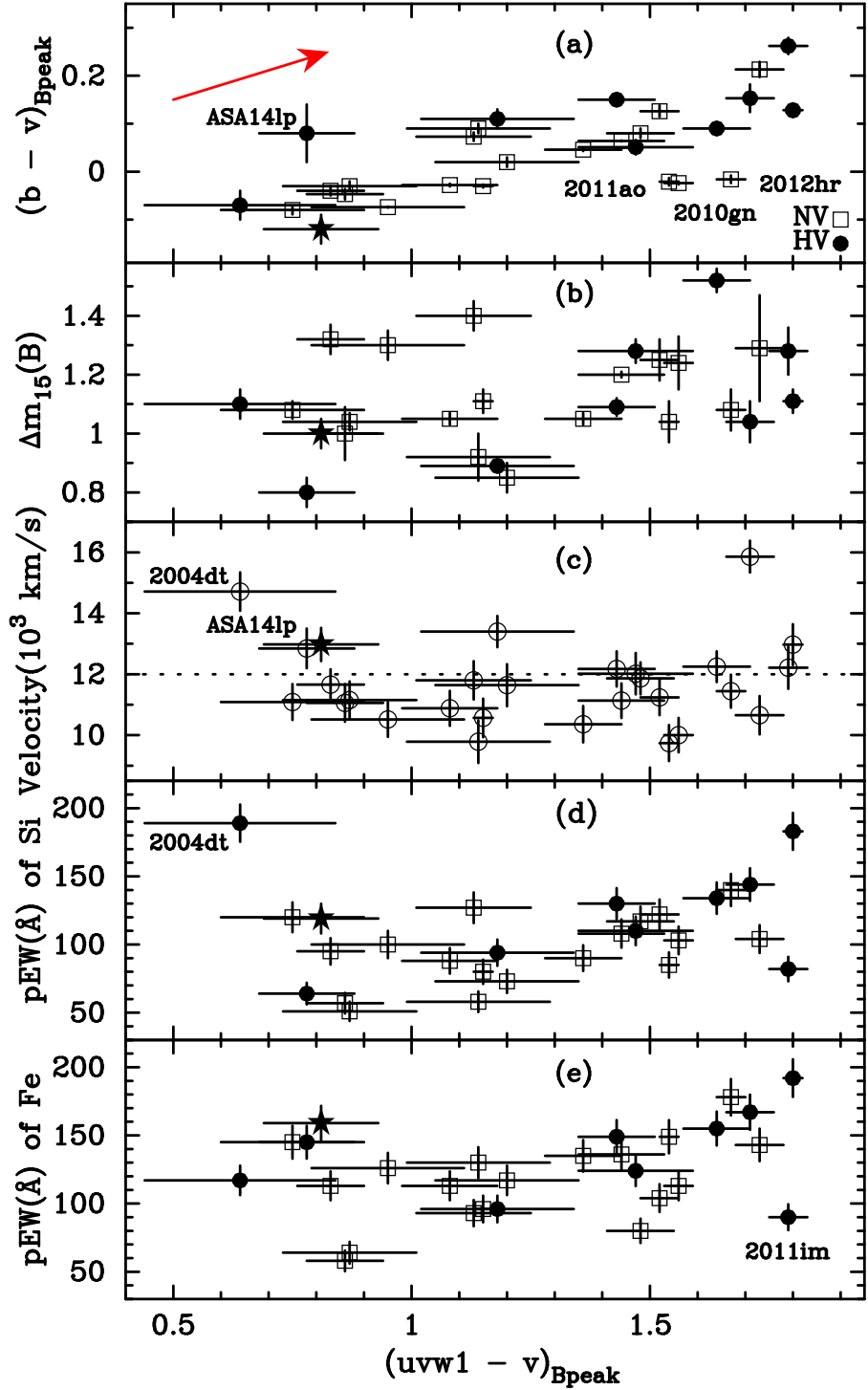}
\caption{a): $uvw1 - v$ vs. $b - v$ color on maximum light. The sample includes two groups: HV (filled circles) and NV (open squares) Ia. The reddening vector from the Milky Way extinction law \citep{car89} with $E(B - V)$ = 0.1 is shown by a red arrow; b): $uvw1 - v$ vs. $\Delta$m$_{15}$(B); c): $uvw1 - v$ vs. the velocity of Si~{\sc ii} $\lambda$6355 around the maximum light, the dot line ($v_{Si}$ = 12000 km s$^{-1}$) in this panel is the boundary between HV and NV Ia; d): $uvw1 - v$ vs. pEW of Si~{\sc ii} $\lambda$6355; e): $uvw1 - v$ vs. pEW of Fe~{\sc ii}/{\sc iii} ($\sim$ 5000{\AA}). The data of SN 2013gs are identified as stars.}
\end{figure}

\clearpage

\begin{longrotatetable}
\begin{deluxetable*}{lccccccccccccc}
\tablecaption{Magnitudes of the photometric standards in the field of SN 2013gs}
\tabletypesize{\scriptsize}
  \tablehead{
Star & $\alpha$ & $\beta$ & $U$ & $B$ & $V$ & $R$ & $I$ & $u$ & $g$ & $r$ & $i$ & $z$\\
  }
\startdata
1  & 09:30:42.094 & 46:24:20.27 &  18.381(026) & 17.234(009) & 15.977(010) &  15.160(046) &   14.486(093) & 19.256(034) & 16.605(011) & 15.467(007) & 14.993(010) & 14.783(003) \\
2  & 09:30:43.044 & 46:18:29.37 &  18.956(036) & 17.960(015) & 16.861(014) &  16.173(046) &   15.501(092) & 19.742(047) & 17.413(020) & 16.415(007) & 15.975(011) & 15.783(008) \\
3  & 09:30:45.212 & 46:19:46.45 &  15.693(009) & 15.313(008) & 14.509(009) &  14.071(028) &   13.668(054) & 16.553(009) & 14.867(009) & 14.291(007) & 14.089(007) & 14.047(005) \\
4  & 09:30:46.963 & 46:22:56.27 &  16.865(013) & 16.602(005) & 15.831(006) &  15.396(028) &   14.992(054) & 17.731(015) & 16.154(009) & 15.606(006) & 15.401(007) & 15.350(003) \\
5  & 09:30:54.447 & 46:25:45.57 &  15.955(010) & 16.154(003) & 15.393(004) &  14.942(025) &   14.668(048) & 16.872(004) & 15.602(009) & 15.201(007) & 15.032(007) & 15.024(004) \\
6  & 09:30:56.043 & 46:25:07.16 &  17.985(021) & 17.899(019) & 17.115(007) &  16.645(030) &   16.207(058) & 18.842(005) & 17.455(007) & 16.905(006) & 16.639(005) & 16.578(007) \\
7  & 09:30:57.038 & 46:18:26.44 &  16.367(011) & 16.273(008) & 15.381(005) &  14.913(027) &   14.607(052) & 17.269(011) & 15.678(007) & 15.141(003) & 14.959(006) & 14.927(004) \\
8  & 09:31:14.747 & 46:18:28.56 &  15.845(010) & 16.016(008) & 15.204(006) &  14.727(028) &   14.396(055) &  \nodata & 15.449(008) & 14.987(006) & 14.790(011) & 14.707(005) \\
9  & 09:31:19.112 & 46:23:47.93 &  19.149(041) & 18.100(010) & 16.892(012) &  16.086(045) &   15.418(092) &  \nodata & 17.509(006) & 16.390(003) & 15.896(007) & 15.697(005) \\
10 & 09:31:20.539 & 46:25:37.45 &  18.080(022) & 17.723(018) & 16.893(014) &  16.342(034) &   15.837(067) & 18.932(013) & 17.248(007) & 16.592(004) & 16.289(003) & 16.178(007) \\
11 & 09:31:28.841 & 46:26:48.42 &  18.557(028) & 18.383(015) & 17.610(022) &  17.161(028) &   16.756(054) & 19.383(039) & 17.904(005) & 17.402(011) & 17.165(005) & 17.091(006) \\
12 & 09:31:29.905 & 46:26:42.18 &  17.083(014) & 16.864(007) & 16.078(010) &  15.630(028) &   15.221(054) & \nodata & 16.399(015) & 15.866(006) & 15.649(007) & 15.575(010) \\
13 & 09:31:33.118 & 46:28:01.39 &  17.516(016) & 16.565(007) & 15.421(007) &  14.642(044) &   13.986(090) & 18.424(026) & 15.982(010) & 14.935(006) & 14.484(006) & 14.269(007) \\
14 & 09:31:37.658 & 46:26:04.88 &  20.590(120) & 19.521(022) & 18.003(012) &  17.034(061) &   16.169(124) & \nodata & 18.723(017) & 17.417(012) & 16.716(008) & 16.405(011) \\
15 & 09:31:38.263 & 46:25:13.13 &  18.784(032) & 18.171(020) & 17.265(005) &  16.664(033) &   16.176(065) & \nodata & 17.677(011) & 16.939(008) & 16.613(002) & 16.494(010) \\
\enddata
\end{deluxetable*}
\end{longrotatetable}

\begin{longrotatetable}
\begin{deluxetable*}{lcrccccccccccr}
\tablecaption{Optical Photometric Observations of SN 2013gs by Ground-based Telescopes}
\tabletypesize{\tiny}
  \tablehead{
  UT Date & MJD & Phase\tablenotemark{a} & $U$ & $B$ &  $V$ & $R$ & $I$ & $u$ & $g$ & $r$ & $i$ & $z$ & Telescope \\
  }
\startdata
2013 Nov 30 &  56626.86 & -12.8 &  17.580(047) &  17.523(042) &  17.326(035) &  17.308(036) &  17.135(037) & \nodata & \nodata & \nodata & \nodata & \nodata & TNT\\
2013 Dec 1  &  56627.86 & -11.8 &  17.079(036) &  17.126(036) &  17.004(037) &  16.996(034) &  16.856(036) & \nodata & \nodata & \nodata & \nodata & \nodata & TNT\\
2013 Dec 2  &  56628.81 & -10.9 &  \nodata &  16.756(036) &  16.687(034) &  16.582(032) &  16.525(035) & \nodata & \nodata & \nodata & \nodata & \nodata & TNT\\
2013 Dec 3  &  56629.69 & -10.0 &  \nodata &  16.553(036) &  16.469(037) &  16.422(032) &  16.305(033) & \nodata & \nodata & \nodata & \nodata & \nodata & TNT\\
2013 Dec 5  &  56631.23 &  -8.5 &   \nodata &  16.226(121) &  16.282(070) &  16.000(115) &  16.105(072) & \nodata & \nodata & \nodata & \nodata & \nodata & Copernico\\
2013 Dec 5  &  56631.80 &  -7.9 &  15.776(024) &  16.076(034) & 16.010(031) &  15.922(032) &  15.854(033) & \nodata & \nodata & \nodata & \nodata & \nodata & TNT\\
2013 Dec 5  &  56632.15 &  -7.5 &  16.022(102) & 15.991(066) & 16.215(104) & 15.993(067) & 16.075(100) & \nodata & \nodata & \nodata & \nodata & \nodata & Copernico\\
2013 Dec 6  &  56632.77 &  -6.9 &  15.597(059) &  15.965(034) &  15.915(030) &  15.754(031) &  15.709(030) & \nodata & \nodata & \nodata & \nodata & \nodata & TNT\\
2013 Dec 7  &  56633.17 &  -6.5 &  15.754(101) & 15.817(053) & 15.954(054) & 15.725(049) & 15.792(048) & \nodata & \nodata & \nodata & \nodata & \nodata & Copernico\\
2013 Dec 8  &  56634.16 &  -5.5 &  15.403(052) & 15.772(074) & 15.809(077) & 15.625(041) & 15.715(056) & \nodata & \nodata & \nodata & \nodata & \nodata & Copernico\\
2013 Dec 8  &  56634.80 &  -4.9 &  15.495(020) &  15.727(034) &  15.645(029) &  15.576(029) &  15.527(029) & \nodata & \nodata & \nodata & \nodata & \nodata & TNT\\
2013 Dec 9  &  56635.04 &  -3.1 &  \nodata  &   15.692(044) & 15.565(046) & \nodata &  \nodata & \nodata & 15.544(063) & 15.609(040) & \nodata & \nodata & LT \\
2013 Dec 9  &  56635.83 &  -3.9 &  15.401(046) &  15.618(043) &  15.521(038) &  15.477(034) &  15.474(033) & \nodata & \nodata & \nodata & \nodata & \nodata & TNT\\
2013 Dec 10 &  56636.87 &  -2.8 &  15.371(018) &  15.589(034) &  15.471(032) &  15.443(029) &  15.502(031) & \nodata & \nodata & \nodata & \nodata & \nodata & TNT\\
2013 Dec 11  &  56637.15 &  -2.6 &  15.601(066) &  15.735(069) &  15.637(053) &  15.430(058) &  15.850(079) & \nodata & \nodata & \nodata & \nodata & \nodata & Copernico\\
2013 Dec 11 &  56637.85 &  -1.9 &  15.366(022) &  15.544(032) &  15.417(033) &  15.387(030) &  15.507(033) & \nodata & \nodata & \nodata & \nodata & \nodata & TNT\\
2013 Dec 11  &  56637.96 &  -1.7 &  15.600(050) &  15.631(043) &  15.498(039) &  15.450(048) &  15.634(065) & \nodata & \nodata & \nodata & \nodata & \nodata & Copernico\\
2013 Dec 11  &  56639.05 &  -0.7 &  15.134(067) &  15.608(057) &  15.667(111) &  15.498(115) &  15.785(192)  & \nodata & \nodata & \nodata & \nodata & \nodata & Copernico\\
2013 Dec 13 &  56639.82 &   0.1 &  15.423(023) &  15.537(032) &  15.365(029) &  15.326(031) &  15.519(033) & \nodata & \nodata & \nodata & \nodata & \nodata & TNT\\
2013 Dec 14 &  56640.81 &   1.1 &  15.429(029) &  15.573(035) &  15.365(033) &  15.322(029) &  15.555(036) & \nodata & \nodata & \nodata & \nodata & \nodata & TNT\\
2013 Dec 15 &  56641.87 &   2.2 &  15.437(070) &  15.593(032) &  15.374(033) &  15.315(031) &  15.578(033) & \nodata & \nodata & \nodata & \nodata & \nodata & TNT\\
2013 Dec 17 & 56643.18 &   3.4 & \nodata &  15.628(020) & 15.389(027) & \nodata & \nodata & 16.517(027) & 15.407(021) & 15.344(019) & 16.001(027) & 16.051(022) & LT\\
2013 Dec 17 &  56643.84 &   4.1 &  \nodata &  15.689(031) &  15.379(010) &  15.321(028) &  15.636(029) & \nodata & \nodata & \nodata & \nodata & \nodata & TNT\\
2013 Dec 18 &  56644.25 &   4.6 & \nodata & 15.684(009) & 15.426(009) & \nodata & \nodata & 16.663(016) & 15.454(010) & 15.385(009) & 16.078(008) & 16.108(010) & LT\\
2013 Dec 18 &  56644.70 &   5.0 &  \nodata &  15.703(033) &  15.393(028) &  15.374(030) &  15.705(032) & \nodata & \nodata & \nodata & \nodata & \nodata & TNT\\
2013 Dec 19 &  56645.27 &   5.6 & \nodata & 15.716(020) & 15.422(009) & \nodata & \nodata & 16.734(028) & 15.522(012) & 15.396(015) & 16.111(012) & 16.140(009)  & LT\\
2013 Dec 19 &  56645.84 &   6.1 &  15.775(025) &  15.761(031) &  15.442(030) &  15.391(031) &  15.720(033) & \nodata & \nodata & \nodata & \nodata & \nodata & TNT\\
2013 Dec 20 &  56646.29 &   6.6 & \nodata & 15.806(019) & 15.466(010) & \nodata & \nodata & 16.872(017) & 15.635(015) & 15.453(008) & 16.139(009) & 16.224(011) & LT\\
2013 Dec 20 &  56646.87 &   7.2 &  \nodata &  15.839(030) &  15.547(030) &  15.503(031) &  15.830(034) & \nodata & \nodata & \nodata & \nodata & \nodata & TNT\\
2013 Dec 21 &  56647.86 &   8.2 &  15.865(094) &  15.944(032) &  15.590(032) &  15.565(032) &  15.894(035) & \nodata & \nodata & \nodata & \nodata & \nodata & TNT\\
2013 Dec 22 &  56648.00 &   8.3 & \nodata & 15.881(053) & 15.517(064) & \nodata & \nodata & 17.044(074) & 15.757(038) & 15.626(046) & 16.277(064) & 16.303(046) & LT\\
2013 Dec 22 &  56648.87 &   9.2 &  \nodata &  15.995(032) &  15.635(032) &  15.637(032) &  15.955(034) & \nodata & \nodata & \nodata & \nodata & \nodata & TNT\\
2013 Dec 23 &  56649.02 &   9.3 &  \nodata &  15.983(020) &  15.592(039) & \nodata &  \nodata & 17.121(036) & 15.807(032) & 15.635(028) & 16.364(026) & 16.336(031) & LT\\
2013 Dec 23 &  56649.86 &  10.2 &  16.078(053) &  16.097(032) &  15.709(030) &  15.743(031) &  16.031(033) & \nodata & \nodata & \nodata & \nodata & \nodata & TNT\\
2013 Dec 24 &  56650.03 &  10.3 &  \nodata &  16.078(032) & 15.638(024) &  \nodata &  \nodata & 17.249(030) & 15.737(025) & 15.730(024) & 16.466(029) & 16.373(020) & LT\\
2013 Dec 25 &  56651.05 &  11.4 & \nodata & 16.193(014) & 15.709(031) & \nodata & \nodata & 17.349(030) & 15.809(016) & 15.772(023) & 16.524(024) & 16.372(030) & LT\\
2013 Dec 25 &  56651.89 &  12.2 &  16.365(031) &  16.260(032) &  15.745(031) &  15.834(031) &  16.079(036) & \nodata & \nodata & \nodata & \nodata & \nodata & TNT\\
2013 Dec 26 &  56652.07 &  12.4 & \nodata & 16.341(033) & 15.777(030) & \nodata & \nodata &  17.448(029) & 15.855(023) & 15.839(033) & 16.586(036) & 16.404(043) & LT\\
2013 Dec 26 &  56652.81 &  13.1 &  16.492(032) &  16.330(034) &  15.814(030) &  15.901(032) &  16.123(037) & \nodata & \nodata & \nodata & \nodata & \nodata & TNT\\
2013 Dec 27 &  56653.09 &  13.4 & \nodata & 16.384(009) & 15.806(008) & \nodata & \nodata & 17.648(025) & 15.939(009) & 15.880(009) & 16.644(017) & 16.393(011) & LT\\
2013 Dec 27 &  56653.69 &  14.0 &  16.563(060) &  16.408(039) &  15.850(033) &  15.957(035) &  16.190(038) & \nodata & \nodata & \nodata & \nodata & \nodata & TNT\\
2013 Dec 28 &  56654.73 &  15.0 &  16.819(035) &  16.552(035) &  15.978(031) &  16.019(032) &  16.159(036) & \nodata & \nodata & \nodata & \nodata & \nodata & TNT\\
2013 Dec 31 &  56657.86 &  18.2 &  17.111(044) &  16.839(040) &  16.167(034) &  16.133(035) &  16.075(037) & \nodata & \nodata & \nodata & \nodata & \nodata & TNT\\
2014 Jan 1  &  56658.87 &  19.2 &  17.249(050) &  16.922(042) &  16.214(035) &  16.139(035) &  15.999(038) & \nodata & \nodata & \nodata & \nodata & \nodata & TNT\\
2014 Jan 4  &  56661.86 &  22.2 &  17.616(068) &  17.245(046) &  16.353(037) &  16.164(037) &  16.014(036) & \nodata & \nodata & \nodata & \nodata & \nodata & TNT\\
2014 Jan 7  &  56664.97 &  25.3 &  18.137(183) & 17.835(120) & 16.494(044) & 16.140(060) & 15.923(077) & \nodata & \nodata & \nodata & \nodata & \nodata & Copernico\\
2014 Jan 8  &  56664.94 &  26.3 &  \nodata & 18.104(154) & 16.591(101) & 16.436(108) & 16.077(151) & \nodata & \nodata & \nodata & \nodata & \nodata & Copernico\\
2014 Jan 10 &  56667.86 &  28.2 &  18.063(054) &  17.669(030) &  16.659(026) &  16.261(026) &  16.011(027) & \nodata & \nodata & \nodata & \nodata & \nodata & LJT\\
2014 Jan 11 & 56668.96 & 29.3 & \nodata & 18.031(074) & 16.583(034) & \nodata & \nodata & 19.209(137) & 17.350(045) & 16.278(029) & 16.401(020) & 16.225(031) & LT\\
2014 Jan 13 & 56670.95 & 30.3 & \nodata & 18.180(030) & 16.752(011) & \nodata & \nodata & 19.228(126) & 17.599(018) & 16.380(020) & 16.507(008) & 16.387(012) & LT\\
2014 Jan 17 &  56674.59 &  34.9 &  \nodata &  \nodata &  17.138(041) &  16.751(039) &  16.240(035) & \nodata & \nodata & \nodata & \nodata & \nodata & TNT\\
2014 Jan 21 &  56678.79 &  39.1 &  \nodata &  18.412(043) &  17.309(042) &  16.997(039) &  16.593(038) & \nodata & \nodata & \nodata & \nodata & \nodata & TNT\\
2014 Jan 26 &  56683.05 &  43.4 &  \nodata & 18.548(136) & 17.472(086) & 17.054(051) & 16.845(043) & \nodata & \nodata & \nodata & \nodata & \nodata & Copernico\\
2014 Jan 26 &  56683.98 & 44.3 & \nodata & 18.507(019) & 17.406(023) & \nodata & \nodata & 20.072(064) & 18.118(025) & 17.083(023) & 17.304(022) & 17.168(020) & LT\\
2014 Jan 27 &  56684.66 & 45.0 &  \nodata &  18.648(045) &  17.548(045) &  17.115(041) &  16.874(041) & \nodata & \nodata & \nodata & \nodata & \nodata & TNT\\
2014 Jan 28 &  56685.02 & 45.3 & \nodata & 18.581(013) & 17.449(009) & \nodata & \nodata & 20.005(051) & 18.147(010) & 17.149(008) & 17.377(010) & 17.209(012) & LT\\
2014 Feb 2  &  56691.85 &  52.1 &  18.957(061) &  18.760(035) &  17.892(034) &  17.461(032) &  17.186(036) & \nodata & \nodata & \nodata & \nodata & \nodata & LJT\\
2014 Feb 7 &   56695.05 &  55.4 & \nodata & 18.745(023) & 17.724(021) & \nodata & \nodata & 20.166(041) & 18.271(029) & 17.457(022) & 17.773(031) & 17.756(026) & LT\\
2014 Feb 9  &  56697.65 &  57.9 &  \nodata &  18.867(124) &  18.028(058) &  17.670(064) &  17.468(070) & \nodata & \nodata & \nodata & \nodata & \nodata & TNT\\
2014 Feb 12 & 56700.94 & 61.2 & \nodata & 18.690(037) & 17.856(015) & \nodata & \nodata & 20.069(112) & 18.279(030) & 17.571(015) & 17.984(015) & 18.034(026) & LT\\
2014 Feb 27 & 56715.52 &  75.8 &  \nodata &  19.078(065) &  18.422(056) &  18.147(035) &  18.166(046) & \nodata & \nodata & \nodata & \nodata & \nodata & TNT\\
2014 Feb 28 & 56716.05 & 76.4 & \nodata & 18.954(029) & 18.270(020) & \nodata & \nodata & 20.579(082) & 18.523(020) & 18.096(026) & 18.563(030) & 18.771(045) & LT\\
2014 Mar 1  &  56717.52 &  77.8 &  \nodata &  19.084(105) &  18.452(048) &  18.176(030) &  18.230(044) & \nodata & \nodata & \nodata & \nodata & \nodata & TNT\\
2014 Mar 9  &  56725.09 &  85.4 &  \nodata & 19.428(162) & 18.603(073) & 18.403(067) & 18.608(072) & \nodata & \nodata & \nodata & \nodata & \nodata & Copernico\\
2014 Apr 2  &  56749.52 & 109.8 &  \nodata &  19.578(150) &  19.305(120) &  19.328(110) &  19.707(120) & \nodata & \nodata & \nodata & \nodata & \nodata & LJT\\
\enddata
\end{deluxetable*}
  \tablenotetext{a}{Relative to the epoch of $B$-band maximum (JD = 2,456,640.2).}
\end{longrotatetable}

\begin{table}
\caption{Unfiltered Photometric Observations of SN 2013gs by 0.6-m Schmidt Telescope.}
{\small
  \begin{tabular}{lcrcc}
  \tableline\tableline
  UT Date & MJD & Phase\tablenotemark{a} & Magnitude & Error\\
  \tableline
2013 Nov 29 &  56625.84 &	-13.86 & 17.26 & 0.18\\
2013 Nov 29 &  56625.90 &	-13.80 & 17.15 & 0.18\\
2013 Dec 2  &  56628.86 &	-10.84 & 16.11 & 0.18\\
2013 Dec 2  &  56628.92 &	-10.78 & 16.12 & 0.18\\
2013 Dec 5  &  56631.84 &	 -7.86 & 15.51 & 0.17\\
2013 Dec 5  &  56631.89 &	 -7.81 & 15.45 & 0.16\\
2013 Dec 8  &  56634.85 &	 -4.85 & 15.15 & 0.21\\
2013 Dec 8  &  56634.90 &	 -4.80 & 15.16 & 0.16\\
2013 Dec 15 &  56641.76 &	  2.06 & 15.02 & 0.18\\
2013 Dec 15 &  56641.82 &	  2.12 & 15.07 & 0.20\\
2013 Dec 23 &  56649.85 &	 10.15 & 15.39 & 0.18\\
2013 Dec 23 &  56649.91 &	 10.21 & 15.37 & 0.22\\
2013 Dec 26 &  56652.89 &	 13.19 & 15.56 & 0.16\\
2014 Jan 2  &  56659.78 &	 20.08 & 15.96 & 0.23\\
2014 Jan 2  &  56659.83 &	 20.13 & 15.92 & 0.18\\
2014 Jan 9  &  56666.83 &	 27.13 & 16.04 & 0.17\\
2014 Jan 9  &  56666.88 &	 27.18 & 16.09 & 0.17\\
2014 Jan 19 &  56676.71 &	 37.01 & 16.64 & 0.20\\
2014 Jan 27 &  56684.68 &	 44.98 & 16.95 & 0.17\\
2014 Jan 27 &  56684.73 &	 45.03 & 16.98 & 0.17\\
2014 Feb 9  &  56697.62 &	 57.92 & 17.22 & 0.18\\
2014 Feb 9  &  56697.72 &	 58.02 & 17.27 & 0.18\\
2014 Feb 19 &  56707.74 &  68.04 & 17.64 & 0.28\\
2014 Feb 24 &  56712.80 &  73.10 & 17.72 & 0.18\\
2014 Feb 28 &  56716.71 &  77.01 & 17.77 & 0.21\\
2014 Mar 15 &  56731.80 &  92.10 & 18.09 & 0.27\\
2014 Mar 21 &  56737.62 &  97.98 & 18.24 & 0.18\\
2014 Mar 24 &  56740.62 & 100.98 & 18.27 & 0.13\\
  \tableline
  \end{tabular}
  \tablenotetext{a}{Relative to the epoch of $B$-band maximum (JD = 2,456,640.2).}
}
\end{table}

\begin{table}
\caption{\emph{Gehrels Swift} UVOT Photometry of SN 2013gs}
{\small
  \begin{tabular}{lcrccccc}
  \tableline\tableline
  UT Date & MJD & Phase\tablenotemark{a} & $u$ & $b$ & $v$ & $uvw1$ & $uvw2$\\
  \tableline
2013 Dec  2 & -11.47 & 56628.23 & 17.339(102) & 16.894(072) & 17.031(143) & 19.184(271) & \nodata \\
2013 Dec  4 & -9.53 & 56630.17 & 16.380(070) & 16.382(062) & 16.382(078) & 18.135(112) & 19.604(239) \\
2013 Dec  6 & -6.97 & 56632.73 & 15.723(064) & 15.909(058) & 16.016(070) & 17.437(085) & 19.063(169) \\
2013 Dec  8 & -5.46 & 56634.24 & 15.520(068) & 15.726(060) & 15.764(077) & 17.168(095) & 18.972(201) \\
2013 Dec 10 & -3.53 & 56636.17 & 15.294(062) & 15.541(055) & 15.607(073) & 17.041(091) & 18.824(184) \\
2013 Dec 12 & -0.86 & 56638.84 & 15.329(061) & 15.462(053) & 15.540(068) & 17.135(086) & 18.589(147) \\
2013 Dec 14 & 0.36 & 56640.06 & 15.503(069) & 15.466(054) & 15.423(070) & 17.082(093) & 18.709(174) \\
2013 Dec 16 & 3.13 & 56642.83 & 15.774(071) & 15.581(056) & 15.419(069) & 17.408(102) & 18.906(190) \\
2013 Dec 18 & 4.73 & 56644.43 & 15.957(074) & 15.680(058) & 15.480(071) & 17.508(108) & 19.081(215) \\
2013 Dec 20 & 6.54 & 56646.24 & 16.260(081) & 15.836(063) & 15.420(071) & 17.759(124) & 19.434(279) \\
2013 Dec 23 & 9.70 & 56649.40 & 16.488(074) & 16.091(061) & 15.682(066) & 18.006(112) & 19.789(288) \\
2013 Dec 26 & 12.90 & 56652.60 & 16.817(080) & 16.458(064) & 15.843(068) & 18.321(129) & 19.894(307) \\
2013 Dec 29 & 16.17 & 56655.87 & 17.250(101) & 16.791(072) & 16.044(078) & 18.867(199) & \nodata \\
2013 Jan  1 & 19.14 & 56658.84 & 17.472(111) & 17.056(078) & 16.054(078) & 19.061(207) & \nodata \\
2013 Jan  4 & 21.77 & 56661.47 & 17.912(135) & 17.307(083) & 16.344(085) & 19.588(319) & \nodata \\
2013 Jan 22 & 40.24 & 56679.94 & 19.002(225) & 18.622(155) & 17.570(153) & \nodata & \nodata \\
2013 Jan 29 & 47.16 & 56686.82 & 19.225(267) & 18.801(175) & 17.624(179) & \nodata & \nodata \\
  \tableline
  \end{tabular}
  \tablenotetext{a}{Relative to the epoch of $B$-band maximum (JD = 2,456,640.2).}
}
\end{table}

\begin{table}
\caption{Light Curve Parameters of SN 2013gs}
{\small
  \begin{tabular}{lcrcccccccc}
  \tableline\tableline
Band & $t_{max}$ & $m_{peak}$ & ${\Delta}m_{15}$ & ${\Delta}m_{35}$ & ${\Delta}m_{60}$ & $\beta$\tablenotemark{a}\\
     & -2,456,000 & (mag) & (mag) & (mag) & (mag) & mag (100 days)$^{-1}$\\  \tableline
unfilter & 641.7 $\pm$ 1.2 & 15.01 $\pm$ 0.05 & 0.66 $\pm$ 0.22 & 1.55 $\pm$ 0.22 & 2.43 $\pm$ 0.20 & 2.37 $\pm$ 0.07\\
$uvw2$   & 639.6 $\pm$ 0.4 & 18.32 $\pm$ 0.05 & \nodata & \nodata & \nodata & \nodata\\
$uvw1$   & 638.6 $\pm$ 0.3 & 16.98 $\pm$ 0.04 & \nodata & \nodata & \nodata & \nodata\\
$U$ & 640.0 $\pm$ 0.3 & 15.36 $\pm$ 0.03 & 1.46 $\pm$ 0.04 & \nodata & \nodata & \nodata\\
$B$ & 640.2 $\pm$ 0.2 & 15.53 $\pm$ 0.02 & 1.00 $\pm$ 0.05 & \nodata & 3.35 $\pm$ 0.13 & 1.39 $\pm$ 0.17\\
$V$ & 642.1 $\pm$ 0.2 & 15.38 $\pm$ 0.02 & 0.60 $\pm$ 0.05 & 1.76 $\pm$ 0.04 & 2.73 $\pm$ 0.06 & 2.70 $\pm$ 0.15\\
$R$ & 640.6 $\pm$ 0.2 & 15.33 $\pm$ 0.03 & 0.67 $\pm$ 0.04 & 1.32 $\pm$ 0.04 & 2.44 $\pm$ 0.06 & 3.34 $\pm$ 0.12\\
$I$ & 640.0 $\pm$ 0.3 & 15.49 $\pm$ 0.03 & 0.67 $\pm$ 0.05 & 0.75 $\pm$ 0.04 & 2.07 $\pm$ 0.07 & 4.38 $\pm$ 0.08\\
  \tableline
  \end{tabular}
  \tablenotetext{a}{The late-time decline rate of light curves.}
}
\end{table}

\begin{table}
\caption{Journal of Spectroscopic Observations of SN 2013gs}
  \begin{tabular}{lcrlcl}
  \tableline\tableline
  UT Date & MJD & Phase\tablenotemark{a} & Range(\AA) & Resolution(\AA)\tablenotemark{b} & Instrument\\  \tableline
2013 Dec 1 & 56627.3  & -12.4 & 3500-8800 & 3 & YNAO 2.4 m YFOSC\\
2013 Dec 3 & 56629.9  &  -9.8 & 3800-8700 & 4 & BAO 2.16 m BFOSC\\
2013 Dec 5 & 56631.2  &  -8.5 & 3700-8200 & 4 & Asiago 1.82 m AFOSC\\
2013 Dec 6 & 56632.1  &  -7.6 & 3300-10000 & 5 & Asiago 1.82 m AFOSC\\
2013 Dec 6 & 56632.8  &  -6.9 & 3600-8700 & 4 & BAO 2.16 m BFOSC\\
2013 Dec 7 & 56633.1  &  -6.6 & 3500-10000 & 5 & Asiago 1.82 m AFOSC\\
2013 Dec 8 & 56634.1  &  -5.6 & 3300-10000 & 5 & Asiago 1.82 m AFOSC\\
2013 Dec 9 & 56635.8  &  -3.9 & 3800-8000 & 4 & BAO 2.16 m OMR\\
2013 Dec 11 & 56637.1  &  -2.6 & 3500-10000 & 5 & Asiago 1.82 m AFOSC\\
2013 Dec 12 & 56638.0  &  -1.7 & 3500-10000 & 5 & Asiago 1.82 m AFOSC\\
2013 Dec 13 & 56639.0  &  -0.7 & 3300-10000 & 5 & Asiago 1.82 m AFOSC\\
2013 Dec 16 & 56642.4  &  +2.7 & 3500-9100 & 3 & YNAO 2.4 m YFOSC\\
2013 Dec 18 & 56644.5  &  +4.8 & 3200-9200 & 2 & FNT 2.0 m FLOYDS\\
2013 Dec 27 & 56653.7  &  +14.0 & 3800-8700 & 4 & BAO 2.16 m BFOSC\\
2013 Dec 27 & 56653.9  &  +14.2 & 3600-10000 & 5 & Asiago 1.82 m AFOSC\\
2014 Jan 4 & 56661.7  &  +22.0 & 3800-8700 & 4 & BAO 2.16 m BFOSC\\
2014 Jan 7 & 56664.9  &  +25.2 & 3500-8200 & 5 & Asiago 1.82 m AFOSC\\
2014 Jan 8 & 56665.9  &  +26.2 & 4700-8200 & 5 & Asiago 1.82 m AFOSC\\
2014 Jan 25 & 56683.0  &  +43.3 & 3600-10000 & 5 & Asiago 1.82 m AFOSC\\
2014 Jan 27 & 56684.7  &  +45.0 & 3800-8000 & 5 & BAO 2.16 m BFOSC\\
2014 Feb 25 & 56713.1  &  +73.4 & 3300-8000 & 5 & Asiago 1.82 m AFOSC\\
2014 Mar 9 & 56725.1  &  +85.4 & 3700-8200 & 5 & BAO 2.16 m BFOSC\\
  \tableline
  \end{tabular}
  \tablenotetext{a}{Relative to the $V$-band maximum (JD = 2,456,640.2).}
  \tablenotetext{b}{Approximate spectral resolution (FWHM intensity).}
\end{table}

\end{document}